\documentclass[12pt]{article}
\usepackage{latexsym}
\usepackage{amsfonts}
\usepackage{epsfig}

\catcode`\@=11

\def\section{\@startsection{section}{1}{\z@}{3.5ex plus 1ex minus
 .2ex}{2.3ex plus .2ex}{\bf}}

\def\thesubsection{\arabic{section}.\arabic{subsection}}
\renewcommand{\subsection}[1]{\addtocounter{subsection}{1}
\vspace{2.5mm}\par\noindent {\it \thesubsection . #1}\par
 \vspace{0.5mm} }
\catcode`\@=12
\def\ii{{\mathrm{i}}}

\mathchardef\varGamma="0100 \mathchardef\varDelta="0101
\mathchardef\varTheta="0102 \mathchardef\varLambda="0103
\mathchardef\varXi="0104 \mathchardef\varPi="0105
\mathchardef\varSigma="0106 \mathchardef\varUpsilon="0107
\mathchardef\varPhi="0108 \mathchardef\varPsi="0109
\mathchardef\varOmega="010A

\def\bfone{\relax{\rm 1\kern-.35em 1}}

\DeclareFontFamily{U}{rsf}{} \DeclareFontShape{U}{rsf}{m}{n}{
  <5> <6> rsfs5 <7> <8> <9> rsfs7 <10-> rsfs10}{}
\DeclareMathAlphabet\Scr{U}{rsf}{m}{n}

\begin{document}
\begin{flushright}
CERN-PH-TH/2010-042
\end{flushright}
\vskip 5mm

  \begin{center}{\LARGE \bf Fake Superpotential for Large and Small Extremal Black Holes}
\vskip 1.5cm
{L. Andrianopoli$^{1}$,  R. D'Auria$^1$, S. Ferrara$^2$ and M. Trigiante$^1$}   \end{center}
 \vskip 3mm
\noindent
{\small
$^1$
Dipartimento di Fisica,
  Politecnico di Torino, Corso Duca degli Abruzzi 24, \\
$~~~$ I-10129
  Turin, Italy and INFN, Sezione di Torino, Italy.
 }

\noindent
{\small
$^2$
Physics Department,Theory Unit, CERN,
CH 1211, Geneva 23, Switzerland,\\
$~~~$ and INFN - Laboratori Nazionali di Frascati,
00044 Frascati, Italy.
 }

\vskip 3mm
 \footnotesize{

 \texttt{laura.andrianopoli@polito.it};

 \texttt{riccardo.dauria@polito.it};

  \texttt{sergio.ferrara@cern.ch};

   \texttt{mario.trigiante@polito.it}.}

\vfill
\vskip 1,5 cm

\begin{abstract}
\noindent We consider the fist order, gradient-flow, description of
the scalar fields coupled to  spherically symmetric, asymptotically
flat black holes in extended supergravities. Using the
identification of the fake superpotential with Hamilton's
characteristic function we clarify some of its general properties,
showing in particular (besides reviewing the issue of its duality
invariance) that $W$ has the properties of a Liapunov's function,
which implies that its extrema (associated with the horizon of
extremal black holes) are asymptotically stable equilibrium points
of the corresponding first order dynamical system (in the sense of
Liapunov). Moreover, we show that the fake superpotential $W$ has,
along the entire radial flow, the same flat directions which exist
at the attractor point. This allows to study properties of the ADM
mass also for small black holes where in fact $W$ has no critical
points at finite distance in moduli space. In particular the $W$
function for small non-BPS black holes can always be computed
analytically, unlike for the large black-hole case.
\end{abstract}

 \vfill\eject

\section{Introduction}
It is well known that in $N$-extended supergravity based on
symmetric coset manifolds $G/H$ the dynamics of extremal
(spherically symmetric, asymptotically flat) black holes is encoded
in a {\em ``fake'' superpotential function} $W$
\cite{1order,1order2,HJ,Ceresole:2009iy,Bossard:2009we,Ceresole:2009vp}\footnote{The
idea of ``fake'' supersymmetry was first introduced in
\cite{fake}.}which, for large black holes, is entirely specified by
the {\em duality orbit} \cite{Bellucci:2006xz} of the dyonic charge
vector $\mathcal{P}=(p^\Lambda,q_\Lambda)$ ($\Lambda = 1,\cdots
,n_V$) and the asymptotic values at radial infinity of the scalars
of the theory: $(\phi_0^r) \in G/H$.

It has not been appreciated enough that many properties of the $W$
function are not only true at the horizon of regular extremal black
holes, defined by critical points $\phi_*$ of $W$:
\begin{eqnarray}
\left.\frac{\partial W}{\partial
\phi^r}\right\vert_{\phi^r=\phi^r_*}=0\,,\label{h}
\end{eqnarray}
where space-time is $AdS_2\times S_2$, but in fact they are valid on
the entire radial flow and in particular at spatial infinity where
space-time is flat. All these properties naturally follow from the
identification of the $W$-function with  Hamilton's characteristic
function \cite{HJ,HJ2} of an autonomous Hamiltonian system
\footnote{See also \cite{DW} for related independent works.}. In
\cite{HJ}, the fact that the radial evolution of scalar fields and
the metric of spherically symmetric, asymptotically flat solutions,
is described by an autonomous Hamiltonian system, was used to show
that the problem of defining a first order description of black
holes in terms of a superpotential $W$ is equivalent to a
Hamilton-Jacobi problem: The $W$ function is a solution to the
Hamilton-Jacobi equation associated with the Hamiltonian system and
defines a system of first order \emph{gradient-flow} equations for
the scalar fields. From the identification of the superpotential $W$
with Hamilton's characteristic function, some important general
properties follow:
\begin{itemize}
\item{$W$ is a positive definite function on the moduli space;}
\item{The derivative of $W$ along the flow, moving from the horizon to radial infinity, is always positive;}
\item{$W$ for extremal solutions is duality invariant.}
\end{itemize}
The latter property was  originally conjectured in \cite{1order2}
and later proven in \cite{HJ}. Eventually, in
\cite{Ceresole:2009iy,Bossard:2009we,Ceresole:2009vp} the explicit
construction of $W$ in terms of duality invariants was
completed.\par As we shall prove, the above properties of $W$, in
the presence of a critical point $\phi_*$ of the first order system,
promote $W$ to a Liapunov's function (see for instance \cite{Liap}),
allowing to make precise statements about the \emph{asymptotic
stability} of $\phi_*$ (namely that $\phi_*$ is not just an
attractive equilibrium point, but also stable), with no need of
computing the Hessian of the potential. The existence of $W$, even
in a neighborhood of the critical point, provides an alternative
(and more powerful) characterization of its attractiveness and
stability properties.\par Of particular interest are orbits of
extremal, large, black holes\footnote{Large black holes are
solutions for which a certain (quartic) duality-invariant expression
of the charge vector $\mathcal{P}$, called $I_4(\mathcal{P})$ does
not vanish. For small black holes, on the other hand,
$I_4(\mathcal{P})=0$. A definition of $I_4(\mathcal{P})$ ,and its
$G$--invariant form for symmetric geometries $G/H$ is summarized in
subsection 2.1} in which the critical points are not isolated. This
feature is related, in the symmetric models, to the existence of
flat directions of the scalar potential $V$ \cite{Ferrara:2007tu}.
We shall prove in full generality, using the duality invariance of
$W$, that $W$ and $V$ have the same symmetry properties, and thus
that they also have the same flat directions. These flat directions,
which have an intrinsic group theoretical characterization in terms
of the stabilizer of the duality orbit of the quantized charges
\cite{Bellucci:2006xz}, are also a feature of the central and matter
charges.
\par
An other consequence of the general properties of  $W$ is that the
functional form of $W(i_n,I_4)$, where $i_n$ are the H-invariant
combinations of the moduli $\phi^r$  and charges $\mathcal{P}$, can also
be calculated for $I_4=0$, in which case the {\em classical} horizon
area vanishes and eq. (\ref{h}) has no solutions (in the interior of the
moduli space). More precisely, for $I_4=0$, eq. (\ref{h}) has a
runaway solution $W=0$ at the boundary of the moduli space where some
$\phi^r\to \infty$ \cite{bfk}.\par
 It is the aim of this note to
further specify general properties of the $W$ function for large and
small black holes, such as their moduli spaces and symmetries.
Moreover, depending on the number  $N$ of supersymmetries, $W(I_4=0)$ can be obtained by a suitable
limit of large black hole solutions (where $I_4\neq 0$), in such a
way that $W$ is always given by a calculable algebraic function of the
$H$-invariants. The way the limit is
performed also allows us to simply understand the interplay of BPS
properties of small black holes versus large solutions.

As a byproduct, our analysis shows that $I_4=0$ black holes in
$\mathcal{N}=8$ are always BPS (distinct in three orbits with
different fractions of supersymmetry), while in $\mathcal{N}=4, 2$
theories $I_4=0$ black holes can be non-BPS. Nevertheless their $W$
function in this case can also be analytically computed from a
non-BPS large black hole with vanishing central charge ($Z_{AB}=0$,
$I_4>0$).

The paper is organized as follows: In section 2 we review and
outline the above mentioned properties of the $W$ superpotential. In
particular we give  the general form of $W$ for extremal solutions
and address the issue of asymptotic stability of the critical points
on $W$, by showing that $W$ can be identified with a Liapunov's
function. We also prove that the potential $V$, the superpotential
$W$, together with the central and matter charges, have the same
flat directions defined by the $G$-orbit of the quantized charges
$\mathcal{P}$. In sections 3, 4, 5 we analyze small black holes for
$\mathcal{N}=8,4$ and $2$ respectively. Three appendices containing
derivations of properties of the $W$-function discussed in the text,
together with other mathematical details, are included.


\section{Some General Properties of the $W$ Function}
Let us review some general properties of the fake superpotential
$W$ associated with $U$-duality orbits of static, extremal,
asymptotically flat black hole solutions in an
 extended  supergravity theory with a symmetric scalar manifold
${\Scr M}_{scal}=\frac{G}{H}$.\par Let us consider an extended
supergravity describing $n$ real scalar fields, spanning the
manifold ${\Scr M}_{scal}$ and $n_V$ vector fields $A_\mu^\Lambda$.
The ansatz for the metric and the vector field strengths
$F_{\mu\nu}^\Lambda$, for the kind of black holes we are
considering, is:
\begin{eqnarray}
d\, s^2&=&-e^{2\,U}\,d\,
t^2+e^{-2\,U}\,\left[\frac{d\tau^2}{\tau^4}\,+
\frac{1}{\tau^2}\,(d\, \theta^2+\sin(\theta)\,d\, \varphi^2)\right]\,,\nonumber\\
\mathbb{F}&=&\left(\matrix{F^\Lambda_{\mu\nu}\cr
G_{\Lambda\,\mu\nu} }\right)\,\frac{dx^\mu\wedge
dx^\nu}{2}=e^{2\,U}\mathbb{C}\cdot\mathcal{M}(\phi^r)\cdot\mathcal{P}\,d\,
t\wedge d\,\tau+\mathcal{P}\,\sin(\theta)\,d\,\theta\wedge
d\,\varphi\,,\label{fg}
\end{eqnarray}
where the coordinate $\tau=-1/r$ runs from $0$, at radial infinity,
to $-\infty$ at the horizon, where $e^{U(\tau)}$ vanishes. The
scalar fields are taken to be functions of $\tau$ only:
$\phi^r=\phi^r(\tau)$. The magnetic field strength
$G_{\Lambda\,\mu\nu}$ in (\ref{fg}) is defined, as usual, as:
$G_{\Lambda\,\mu\nu}\propto \, \epsilon_{\mu\nu\rho\sigma}\,\delta
\mathcal{L}/\delta F^{\Lambda}_{\rho\sigma}$, $\mathcal{L}$ begin
the Lagrangian of the theory. The last equation in (\ref{fg}) is
written in a manifestly symplectic covariant form, namely as an
equality between two $2\,n_V$ dimensional symplectic vectors, where
$\mathbb{C}_{MN}$, $M,\,N=1,\dots, 2\,n_V$ is the ${\rm
Sp}(2\,n_V,\mathbb{R})$-invariant matrix:
\begin{eqnarray}
\mathbb{C}&=&\left(\matrix{{\bf 0} & -\bfone \cr \bfone &{\bf 0}
}\right)\,.
\end{eqnarray}
The vector $\mathcal{P}\equiv (p^\Lambda,q_\Lambda)$ consists of the
quantized electric and magnetic charges. Finally the $2\,n_V\times
2\,n_V$ symmetric, negative defined, symplectic matrix
$\mathcal{M}(\phi^r)_{MN}\equiv -(\mathbb{L}\,\mathbb{L}^T)_{MN}$,
$\mathbb{L}(\phi^r)$ being the ${\Scr M}_{scal}$ coset
representative in the fundamental of ${\rm Sp}(2\,n_V,\mathbb{R})$,
can also be written in the familiar form \cite{bhreviews}:
\begin{eqnarray}
\mathcal{M}(\phi^r)&=&\left(\matrix{I+R\,I^{-1}\,R& -R\,I^{-1}\cr
-I^{-1}\,R & I^{-1} }\right)\,,
\end{eqnarray}
where $I_{\Lambda\Sigma}\equiv {\rm
Im}(\mathcal{N})_{\Lambda\Sigma}<0$ is the vector kinetic matrix
while $R_{\Lambda\Sigma}\equiv {\rm
Re}(\mathcal{N})_{\Lambda\Sigma}$ defines the generalized
theta-term. \par Once the electric and magnetic charges of the
solution are assigned, the radial evolution of the $n+1$ fields
$U(\tau),\,\phi^r(\tau)$ is described by the effective action:
\begin{eqnarray}
S_{eff}&=&\int\mathcal{L}_{eff}\,d\tau=\int\left(\dot{U}^2+\frac{1}{2}\,G_{rs}(\phi)\,\dot{
\phi}^r\,\dot{
\phi}^s+e^{2\,U}\,V(\phi,\mathcal{P})\right)\,d\tau\,,\label{lag}
\end{eqnarray}
 together with the Hamiltonian constraint, representing the
extremality condition\footnote{For non extremal black holes the
value of the Hamiltonian on a solution coincides with the square of
the extremality parameter. Notice that the Hamiltonian is not positive definite, being
expressed as the difference of a ``kinetic'' and a positive
``potential'' term (this is in turn due to the fact that the role of
the time variable is played by a spatial coordinate $\tau$). As a
consequence of this we can have non-trivial solutions on which the
Hamiltonian vanishes. These correspond to the extremal black
holes.}:
\begin{eqnarray}
\mathcal{H}_{eff}&=&\dot{U}^2+\frac{1}{2}\,G_{rs}(\phi)\,\dot{ \phi}^r\,\dot{
\phi}^s-e^{2\,U}\,V(\phi,\mathcal{P})=0\,,\label{ham}
\end{eqnarray}
the effective potential being given by $V(\phi,\mathcal{P})\equiv
- \frac{1}{2}\,\mathcal{P}^T\,\mathcal{M}(\phi)\,\mathcal{P}>0$ and
the dot represents the derivative with respect to $\tau$. The radial
evolution of the $n+1$ fields $U(\tau),\,\phi^r(\tau)$ in the
solution admits a first order description \cite{1order,1order2,HJ}
in terms of  a fake superpotential $W(\phi,\mathcal{P})$:
\begin{eqnarray}
\dot{U}&=&e^U\,W\,\,\,\,,\,\,\,\,\,\dot{\phi}^r=2\,e^U\,G^{rs}\,\frac{\partial
W }{\partial\phi^s}\,.\label{1or}
\end{eqnarray}
 If we interpret the fields $U(\tau),\,\phi^r(\tau)$ as
 coordinates of a Hamiltonian system in which the radial variable plays the role of time,
 the first order description (\ref{1or}) is equivalent to solving
 the Hamilton-Jacobi problem with
Hamilton's characteristic function
\begin{equation}
\mathcal{W}(U,\phi)\equiv
2\,e^U\,W(\phi)\,.
\end{equation}
 Indeed, in terms of $W(\phi,\mathcal{P})$, the
Hamilton-Jacobi equation has the form:
\begin{eqnarray}
W^2+2\,G^{rs}\,\frac{\partial W }{\partial\phi^r}\,\frac{\partial
W }{\partial\phi^s}&=&V\,,\label{hj}
\end{eqnarray}
which can also be derived from the Hamiltonian constraint
(\ref{ham}) using (\ref{1or})\footnote{In the case of non-extremal
solutions the Hamilton-Jacobi equation reads $\left(\frac{\partial
\mathcal{W}}{\partial U}\right)^2+2\,G^{rs}(\phi)\,\frac{\partial
\mathcal{W}}{\partial \phi^r}\,\frac{\partial
\mathcal{W}}{\partial \phi^s}=4\,e^{2U}\,V+4\,c^2$, and the
corresponding first order equations have the form
$\dot{U}=\frac{1}{2}\,\frac{\partial \mathcal{W}}{\partial
U},\,\dot{\phi}^r=G^{rs}(\phi)\,\frac{\partial
\mathcal{W}}{\partial \phi^s}$. If $c\neq 0$ however, as it is
apparent from the Hamilton-Jacobi equation, the dynamical system
can have no equilibrium point $\frac{\partial
\mathcal{W}}{\partial U}=\frac{\partial \mathcal{W}}{\partial
\phi^r}=0$.}. We are not interested here in the most general
solution to (\ref{hj}), nor to address the issue of its existence
(see \cite{HJ} for a discussion on this point). We are interested,
instead, in  the $W$ functions associated with classes of extremal
solutions whose general properties are in principle known. They
are completely characterized by the set of quantized charges
$\mathcal{P}$ and the values of the fields at radial infinity:
\begin{eqnarray}
U(\tau=0)&=&0\,\,\,,\,\,\,\,\phi^r(\tau=0)=\phi^r_0\,.
\end{eqnarray}
We shall therefore  simply denote them  by: $U=U(\tau;\phi_0)$ and
$\phi^r=\phi^r(\tau;\phi_0)$. The ADM mass and the scalar charges
at infinity are given by:
\begin{eqnarray}
M_{ADM}(\phi_0,\mathcal{P})&=&\dot{U}(\tau=0)=W(\phi_0,\mathcal{P})\,,\nonumber\\
\Sigma^r(\phi_0,\mathcal{P})&=&
\dot{\phi}^r(\tau=0)=2\,G^{rs}(\phi_0)\,\frac{\partial W
}{\partial\phi^r}(\phi_0,\mathcal{P})\,.\label{admm}
\end{eqnarray}
Regular (\emph{large}) extremal black holes have finite horizon area
$A_H$ and thus near the horizon ($\tau\rightarrow -\infty$) $e^U$
has the following behavior: $e^{-2U}\sim \frac{A_H}{4\pi}\,\tau^2$,
where $A_H=A_H(\mathcal{P})$ is a function of the quantized charges
only. In fact $\mathcal{P}$ transforms under duality (see subsection
2.2) in a symplectic representation of $G$ and $A_H$, as a function
of $\mathcal{P}$, is expressed in terms of the quartic invariant of
$G$ in this representation:
$A_H(\mathcal{P})=4\,\pi\,\sqrt{|I_4(\mathcal{P})|}$ (here we use
the units $c=\hbar=G=1$,  so that the Plank length is one.). Using
eq.s (\ref{1or}) we see that $W$ computed on the solution evolves,
in the near horizon limit, towards
$\sqrt{\frac{A_H}{4\pi}}$.\\
 \noindent As far as the scalar fields
are concerned, due to the attractor mechanism some of them are fixed
at the horizon to values which are totally determined in terms of the
quantized charges, while other scalar fields, which are flat
directions of the potential, are not.
That is, in the presence of flat directions, in the near
horizon limit $\tau\rightarrow -\infty$ the non flat scalars  evolve towards values
which are totally fixed in terms of quantized charges, while the flat directions still depend, in general, on the boundary values  $\phi^r_0$ taken at radial infinity ($\tau=0$).
Since
 only the scalars parametrizing the flat directions may depend
 at the horizon on $\phi^r_0$, the near horizon geometry, which is determined in terms of the potential, will
 only depend on the quantized charges, consistently with the
  attractor mechanism. Summarizing, for large black holes, we have:
\begin{eqnarray}
\lim_{\tau\rightarrow
-\infty}e^{-2U}&=&\sqrt{|I_4(\mathcal{P})|}\,\tau^2
\,\,\,,\,\,\,\,\lim_{\tau\rightarrow
-\infty}\phi^r(\tau)=\phi^r_*\,,\nonumber\\
\lim_{\tau\rightarrow
-\infty}W^2(\phi(\tau;\phi_0),\mathcal{P})&=&W^2(\phi_*,\mathcal{P})=V(\phi_*,\mathcal{P})=\sqrt{|I_4(\mathcal{P})|}\,.\nonumber
\end{eqnarray}
Small black holes are characterized by vanishing horizon area, i.e.
by quantized charges for which $I_4(\mathcal{P})=0$. For
$\tau\rightarrow -\infty$ the warp factor has the following
behavior: $e^{-2U}\sim \tau^\alpha$, $\alpha<2$. In the same limit
scalar fields typically flow to values which are at the boundary of
the scalar manifold. Either for large or for small solutions, from
the first of (\ref{1or}) we deduce the following boundary condition
for $W$:
 \begin{eqnarray}
\lim_{\tau\rightarrow
-\infty}e^{U(\tau;\phi_0)}\,W(\phi(\tau;\phi_0),\mathcal{P})&=&\lim_{\tau\rightarrow
-\infty}\dot{U}=0\,.
\end{eqnarray}
This allows us to write $W(\phi,\mathcal{P})$ for the two kinds of
solutions  in the following form (see \cite{HJ}):
\begin{eqnarray}
W(\phi_0,\mathcal{P})&=&\int_{-\infty}^0
e^{2\,U(\tau;\phi_0)}\,V(\phi(\tau;\phi_0),\mathcal{P})\,d\tau\,.\label{Wgen}
\end{eqnarray}
It should be stressed that the above expression allows to write
the $W$ function for a given class of solutions as  a free
function of the point $\phi_0$ on the scalar manifold and of the
quantized charges: Given a charge vector $\mathcal{P}$ and a point
$\phi_0=(\phi_0^r)$ in ${\Scr M}_{scal}$, the corresponding value
of $W$ is given by the integral over $\tau$ of $e^{2U}\,V$,
computed along the unique solution originating at infinity in
$\phi_0$.
\subsection{$I_4$ Invariant for $\mathcal{N}=2,4,8$
Supergravities}\label{i4} In $\mathcal{N}>2$ theories and in
$\mathcal{N}=2$ theories based on symmetric spaces for the (vector
multiplet) scalar fields, the entropy \emph{area law} reads (the
Boltzmann constant $k_B$ being one in our units):
\begin{eqnarray}
S&=&\frac{A_H}{4}=\pi\,\sqrt{|I_4(\mathcal{P})|}\,,
\end{eqnarray}
where, as anticipated in the previous section, $I_4(\mathcal{P})$ is
a certain quartic invariant of the dyonic charge vector
$\mathcal{P}$ and depends on the particular theory under
consideration. Since $I_4(\mathcal{P})$ is moduli-independent, it
can be expressed either in terms of the quantized charges
$\mathcal{P}$ or in terms of the (dressed) central and matter
charges $Z_{AB}(\phi,\,\mathcal{P}),\,Z_{I}(\phi,\,\mathcal{P})$
(see subsection 2.3 for a precise definition of the latter). For our
convenience we recall here the actual form of $I_4(\mathcal{P})$ in
terms of the central and matter charges.
\par
For $\mathcal{N}=2$
theories, based on special geometry, we can define five
$H$--invariant quantities $i_n$, as follows \cite{Ceresole:2009iy}:
\begin{eqnarray}
i_1&\equiv & Z\,\overline{Z}\,,\nonumber\\
i_2&\equiv & g^{i\bar{\jmath}} Z_i\,\overline{Z}_{\bar{\jmath}}\,,\nonumber\\
i_3&\equiv &\frac{1}{3}\,{\rm
Re}\left(Z\,N_3(\overline{Z}_{\bar{\imath}})\right)\,,\nonumber\\
i_4&\equiv &-\frac{1}{3}\,{\rm
Im}\left(Z\,N_3(\overline{Z}_{\bar{\imath}})\right)\,,\nonumber\\
i_5&\equiv &
g^{i\bar{\imath}}\,C_{ijk}\,\overline{C}_{\bar{\imath}\bar{\jmath}\bar{k}}\,
\overline{Z}^j\,\overline{Z}^k\,{Z}^{\bar{\jmath}}\,{Z}^{\bar{k}}\,,\nonumber
\end{eqnarray}
where $N_3(\overline{Z}_{\bar{\imath}})\equiv
C_{ijk}\,\overline{Z}^i\,\overline{Z}^j\,\overline{Z}^k$, $Z_i\equiv
D_iZ$ and $Z^{\bar{\imath}}\equiv g^{\bar{\imath} i}\,Z_i$. In terms
of these quantities the quartic invariant reads:
\begin{eqnarray}
I_4&=&(i_1-i_2)^2+4\,i_4-i_5=I_4(\mathcal{P})\,,
\end{eqnarray}
where, as anticipated in the previous subsection, $\mathcal{P}$
transforms in a symplectic representation of $G$ and
$I_4(\mathcal{P})$ is the only non-vanishing invariant quantity
built out of the charge vector. Note that, for the quadratic series
($C_{ijk}=0$) we have: $I_4=I_2^2$, where $I_2\equiv |i_1-i_2|$.\par
For $\mathcal{N}=4$, we can define two $\rm SU(4)\times SO(n)$
invariants:
\begin{eqnarray}
S_1&\equiv &
\frac{1}{2}\,Z_{AB}\,\overline{Z}^{AB}-Z_{{I}}\,\overline{Z}_{\bar{J}}\,\delta^{I \bar{J}}\,,\nonumber\\
S_2&\equiv &
\frac{1}{4}\,\epsilon^{ABCD}\,Z_{AB}\,Z_{CD}-Z_{{I}}\,{Z}_J\,\delta^{IJ}\,,\nonumber
\end{eqnarray}
in terms of the central charges $Z_{AB}=-Z_{BA}$, $A,B=1,\dots, 4$,
and the $n$ matter charges $Z_I$, $I=1,\dots, n$. Then the unique
quartic $G=\rm SL(2,\mathbb{R})\times SO(6,n)$-invariant reads:
\begin{eqnarray}
I_4^{(\mathcal{N}=4)}(\mathcal{P})&\equiv &
S_1^2-|S_2|^2\,,\label{I44}
\end{eqnarray}
and the black hole potential is:
\begin{eqnarray}
V^{(\mathcal{N}=4)}(\phi,\,\mathcal{P})&=&
\frac{1}{2}\,Z_{AB}\,\overline{Z}^{AB}+Z_{{I}}\,\overline{Z}^I\,.
\end{eqnarray}\par
Finally, in the $\mathcal{N}=8$ theory the Cartan $G=\rm
E_{7(7)}$-quartic invariant is given by the expression \cite{cj}:
\begin{eqnarray}
I_4^{(\mathcal{N}=8)}(\mathcal{P})&\equiv & {\rm
Tr}[(\mathbb{Z}\,\mathbb{Z}^\dagger)^2]-[{\rm
Tr}(\mathbb{Z}\,\mathbb{Z}^\dagger)]^2+8\,{\rm
Re}[Pf(\mathbb{Z})]\,,
\end{eqnarray}
where $\mathbb{Z}\equiv (Z_{AB})=-\mathbb{Z}^T$, $A,B=1,\dots, 8$,
is the complex central charge matrix \cite{kalkol}. In terms of the
four skew-eigenvalues $z_i$, $i=1,\dots, 4$, of $Z_{AB}$,
$I_4^{(\mathcal{N}=8)}$ reads:
\begin{eqnarray}
I_4^{(\mathcal{N}=8)}(\mathcal{P})&\equiv &
\sum_{i=1}^4|z_i|^4-2\,\sum_{i<j}|z_i|^2\,|z_j|^2+4\,(z_1z_2z_3z_4+\bar{z}_1\,\bar{z}_2\,\bar{z}_3\,\bar{z}_4)\,.
\end{eqnarray}
The black hole effective potential has the following form:
\begin{eqnarray}
V^{(\mathcal{N}=8)}(\phi,\,\mathcal{P})&=&
\frac{1}{2}\,Z_{AB}\,\overline{Z}^{AB}=\sum_{i=1}^4|z_i|^2\,.
\end{eqnarray}
In any extended supergravity, BPS solutions are described by
$W=|z_h|$, where $z_h$ is the highest skew-eigenvalue (i.e.
eigenvalue with highest modulus) of the central charge matrix
$Z_{AB}$ (for $\mathcal{N}=2$, $Z_{AB}=Z\,\epsilon_{AB}$, $A,B=1,2$,
and $z_h=Z$). Therefore it is also true that:
\begin{eqnarray}
V&=&|z_h|^2+2\,G^{rs}\,\partial_r|z_h|\,\partial_s|z_h|\,.
\end{eqnarray}
If however $\mathcal{P}$ is not in a BPS orbit, the flow defined by
$W=|z_h|$ does not correspond to a physically acceptable solution
and a different $W$-function should be used.
\par
In particular, in the $\mathcal{N}=8$ case for non-BPS
configurations  the corresponding $W$-function satisfies the
following inequalities:
\begin{eqnarray}
&&|z_h|^2<W^2\le 4\,|z_h|^2\,,
\end{eqnarray}
the lower bound being saturated only for BPS solutions. The upper
bound originates from the general property: $W^2\le V\le
4\,|z_h|^2$. For non-BPS large black holes, it can be proven that,
at the attractor point, $|z_i|=\rho=|z_h|$ and the upper bound is
saturated: $W=2\,\rho$.
\subsection{The $W$ Function and Duality}\label{duality}
It is known that the on-shell global symmetries of an extended
supergravity, at the classical level, are encoded in the isometry
group $G$ of the scalar manifold (if non-empty), whose action on the
scalar fields is associated with a simultaneous linear symplectic
action on the field strengths $F^\Lambda$ and their duals
$G_\Lambda$. This duality action of $G$ is defined by a symplectic
representation $D$ of $G$:
\begin{eqnarray}
g\in G&:&\cases{\phi^r\rightarrow \phi^{r\,\prime}=g\star
\phi^r\cr \left(\matrix{F^\Lambda\cr G_\Lambda}\right)\rightarrow
D(g)\cdot \left(\matrix{F^\Lambda\cr G_\Lambda}\right) }\,,
\end{eqnarray}
where $g\star$ denotes the non-linear action of $g$ on the scalar
fields and $D(g)$ is the $2\,n_v\times 2\,n_v$ symplectic matrix
associated with $g$. The matrix $\mathcal{M}(\phi)$ transforms
under $G$ as follows:
\begin{eqnarray}
\mathcal{M}(g\star\phi)&=&D(g)^{-T}\,\mathcal{M}(\phi)\,D(g)^{-1}\,.\label{mtrans}
\end{eqnarray}
A duality transformation $g\in G$ maps a black hole solution
$U(\tau),\phi^r(\tau)$ with charges $\mathcal{P}$ into a new
solution $U^\prime(\tau)=U(\tau),\,\phi^{\prime\,r}(\tau)=g\star
\phi^r(\tau)$ with charges $\mathcal{P}^\prime=D(g)\,\mathcal{P}$.
More specifically, if $U(\tau),\phi^r(\tau)$ is defined by the
boundary condition $\phi_0$ for the scalar fields,
$U^\prime(\tau)=U(\tau),\,\phi^{\prime\,r}(\tau)$ is the
\emph{unique solution}, within our class,  with charges
$\mathcal{P}^\prime$ defined by the boundary condition
$\phi_0^\prime=g\star\phi_0$
\begin{eqnarray}
g\in G&:&\cases{U(\tau;\,\phi_0)\cr\phi(\tau;\,\phi_0)\cr
\mathcal{P}}\,\,\,\longrightarrow\,\,
\,\,\,\cases{U^\prime(\tau;\,g\star\phi_0)=
U(\tau;\,\phi_0)\cr\phi^\prime(\tau;\,g\star\phi_0)=g\star\phi(\tau;\,\phi_0)
\cr \mathcal{P}^\prime=D(g)\,\mathcal{P}}\,.\label{utrans}
\end{eqnarray}
Using eq.s (\ref{mtrans}) and (\ref{utrans}), we see that the
effective potential is invariant if we act on $\phi^r$ and
$\mathcal{P}$ by means of $G$ simultaneously:
\begin{eqnarray}
V(\phi,\mathcal{P})&=&V(g\star\phi,D(g)\,\mathcal{P})\,.\label{hinv}
\end{eqnarray}
 This implies that $V$, as a function of the scalar fields and quantized charges, is
$G$-invariant. From this property of $V$ it follows  that the
effective action (\ref{lag}) and the extremality constraint
(\ref{ham}) are manifestly duality invariant. Let us show now that
the $W$ function shares with $V$ the same symmetry property
(\ref{hinv}), namely that it is $G$-invariant as well:
\begin{eqnarray}
W(\phi,\mathcal{P})&=&W(g\star\phi,D(g)\,\mathcal{P})\,.\label{hinvw}
\end{eqnarray}
This is easily shown using the general form (\ref{Wgen}) and eq.s
(\ref{utrans}):
\begin{eqnarray}
W(g\star\phi_0,\,D(g)\,\mathcal{P})&=&
\int_{-\infty}^0e^{2\,U^\prime(\tau;\,g\star\phi_0)}\,V(\phi^\prime(\tau;g\star\phi_0),D(g)\,\mathcal{P})\,d\tau=\nonumber\\&=&
\int_{-\infty}^0e^{2\,U(\tau;\,\phi_0)}\,V(g\star\phi(\tau;\phi_0),D(g)\,\mathcal{P})\,d\tau=\nonumber\\&=&
\int_{-\infty}^0e^{2\,U(\tau;\,\phi_0)}\,V(\phi(\tau;\phi_0),\mathcal{P})\,d\tau=W(\phi_0,\mathcal{P})\,.\label{hinvw2}
\end{eqnarray}
Being the ADM mass expressed in terms of $W$, see eq. (\ref{admm}),
it is a $G$-invariant quantity as well:\begin{eqnarray}
M_{ADM}(\phi_0,\mathcal{P})&=&M_{ADM}(g\star\phi_0,D(g)\,\mathcal{P})\,.\label{hinvadm}
\end{eqnarray}
Extremal black-holes can be grouped into orbits with respect to the
duality action (\ref{utrans}) of $G$. These orbits are characterized
in terms of $G$-invariant functions of the scalar fields and the
quantized charges, which are expressed in terms of $H$-invariant
functions of the central and matter charges. One of these is the
scalar-independent quartic invariant $I_4(\mathcal{P})$ of $G$ which
defines the area of the horizon for large black holes. Small black
holes, on the other hand, belong to the orbits in which
$I_4(\mathcal{P})=0$.
\subsection{The Issue of Stability: Asymptotic Stability of the
Critical Points} Let us notice, from eq. (\ref{Wgen}), that $W$ is
always positive definite, since the effective potential is. Moreover
its derivative along the solution $\phi^r(\tau)$ is positive
definite  as well (except in $\phi_*$ where it vanishes):
\begin{eqnarray}
\frac{dW}{d\tau}&=&\dot{\phi}^r\partial_r
W=e^{-U}\,G_{rs}(\phi)\,\dot{\phi}^r\,\dot{\phi}^s>0\,.
\end{eqnarray}
We see that, if $\phi_*$ is isolated, $W$ has the properties of a
Liapunov's function and thus, in virtue  of Liapunov's theorem,
$\phi_*$ is a \emph{stable attractor point} (we refer the reader to
Appendix \ref{liapunov} for a brief review of the notion of
asymptotic stability in the sense of Liapunov and of Liapunov's
theorem, see also standard books like \cite{Liap}). This conclusion
extends to models based on a generic (not necessarily homogeneous)
scalar manifold:
 The very existence of the $W$-function (i.e. of a solution to the Hamilton-Jacobi equation) even just in
 a neighborhood of an isolated critical point $\phi_*$ is enough to guarantee asymptotic stability of $\phi_*$,
  and thus that the horizon is a stable attractor. Let us emphasize that in this case we need not evaluate
 the  Hessian of the potential on $\phi_*$.  In other words the (local) existence of $W$
can be taken as an alternative and more powerful characterization of
the attractiveness and stability properties of the horizon point
$\phi_*$.
\par There is a class of large extremal solutions, however, in which
the critical points, defining the near-horizon behavior of the
scalar fields, are not isolated but rather span a hypersurface
$\mathcal{C}$ of the scalar manifold. This is the case of the
non-BPS solutions with $I_4<0$ in the symmetric models. As we are
going to show below, in full generality, the existence of this locus
of critical points is related to the existence of $n_f<n$ \emph{flat
directions} $\varphi^\alpha$, $\alpha =1,\dots, n_f$, of
\emph{both} the scalar potential $V$ and the $W$ function. The
critical hypersurface $\mathcal{C}$ has in this case dimension $n_f$
and is spanned by $(\varphi^\alpha )$. As far as the global behavior of
the flows is concerned, the analysis of the simple STU model  (see
\cite{Gimon:2007mh} for a discussion on this point) suggests a
general property: The scalar manifold can be decomposed in
hypersurfaces ${\Scr M}_{(\alpha)}$ of dimension $n-n_f$ which intersect
the hypersurface of critical points $\mathcal{C}$ in a single point
$\phi_*\vert_{\alpha}$ characterized by fixed values $\varphi^\alpha$ of the
flat directions. The hypersurfaces ${\Scr M}_{(\alpha)}$ have the
property of being \emph{invariant} with respect to the flow, namely
that, choosing the initial point $\phi_0$ on a given ${\Scr
M}_\alpha$, the entire flow will be contained within the same
hypersurface. Within each ${\Scr M}_{(\alpha)}$ the critical point
$\phi_*\vert_{\alpha}$ is isolated and Liapunov's theorem applies,
implying it is asymptotically stable or, equivalently,  a stable
attractor.
\subsection{The Issue of Flat Directions}\label{ifd}
Let us denote by $G_0\subset G$ the \emph{little group} (or
stabilizer) of the orbit of the quantized charges $\mathcal{P}$
under the action of $G$ \cite{Bellucci:2006xz,Cerchiai:2009pi}:
\begin{eqnarray}
g_0\in G_0&:&\,\,\,D(g_0)\,\mathcal{P}=\mathcal{P}\,. \label{little}
\end{eqnarray}
Of course the embedding of $G_0$ within $G$ depends in general on
$\mathcal{P}$.
 Let us show that the scalar fields $\varphi^\alpha$
spanning the submanifold $G_0/H_0$, $H_0$ being the maximal compact
subgroup of $G_0$, are \emph{flat directions} of the potential and
of the $W$-function, namely that neither $V$  nor $W$, depend on
$\varphi^\alpha$. Since we are interested in the part of the
little group which has a free action on the moduli, we shall define  $G_0$
modulo compact group-factors. For instance  if the little group is ${\rm SU}(3)\times {\rm SU}(2,1)$, we define $G_0$ to be
 ${\rm SU}(2,1)$ and thus $H_0={\rm U}(2)$. For a summary of the orbits of regular extremal
  black holes in the various theories and of the corresponding moduli spaces $G_0/H_0$ see Table 1.
 \par To prove that  $\varphi^\alpha$ are flat directions of both $V$ and $W$,
 let us decompose the $n$ scalar fields
$\phi^r$ into the $\varphi^\alpha $ scalars  parametrizing the submanifold
$G_0/H_0$ and scalars $\varphi^k$, which can be
chosen to transform linearly with respect to $H_0$. Let us stress at
this point that the coordinates $\varphi^\alpha , \varphi^k$ will in
general depend on the original ones $\phi^r$ and on the electric and
magnetic charges, namely:
\begin{eqnarray}
\varphi^\alpha
 &=&\varphi^\alpha
(\phi^r,p^\Lambda,q_\Lambda)\,\,,\,\,\,\,{\varphi}^k
={\varphi}^k(\phi^r,p^\Lambda,q_\Lambda)\,.
\end{eqnarray}
Let us choose, for convenience, a basis of coordinates  in the moduli space such that the first $n_f$ components of $\phi^r$ coincide with the $\varphi^\alpha$,  the others being $\varphi^k$, that is $\phi^\alpha=\varphi^\alpha$, $\phi^k=\varphi^k$.
 We can move along the $\phi^\alpha $ direction
through the action of isometries in $G_0$. We shall consider
infinitesimal isometries in $G_0$ whose effect is to shift the
$\alpha$-scalars only:
\begin{eqnarray}
g_0\in G_0\,\,&:&\,\,\,
\phi^r\,\,\rightarrow\,\,(g_0\star\phi)^r=\phi^r+\delta^r_\alpha \,\delta\phi^\alpha
\,\,\,\,,\,\,\,\,\,\mathcal{P}\,\rightarrow\,\,\mathcal{P}^\prime=\mathcal{P}+\delta
\mathcal{P}=\mathcal{P}\,,
\end{eqnarray}
where we have used the definition of $G_0$, (\ref{little}). Let us now use eq.s
(\ref{hinv}) and (\ref{hinvw}) to evaluate the corresponding
infinitesimal variations of $V$ and $W$:
\begin{eqnarray}
V(\phi^r,\mathcal{P})&=&V(\phi^r+\delta \phi^r,\mathcal{P}+\delta
\mathcal{P})=V(\phi^k,\phi^\alpha+\delta \phi^\alpha,\mathcal{P})\,.\nonumber\\
W(\phi^r,\mathcal{P})&=&W(\phi^r+\delta \phi^r,\mathcal{P}+\delta
\mathcal{P})=W(\phi^k,\phi^\alpha+\delta \phi^\alpha,\mathcal{P})\,.
\end{eqnarray}
We conclude that $\frac{\partial V}{\partial \phi^\alpha
}=\frac{\partial W}{\partial\phi^\alpha }=0$, namely that $\phi^\alpha $
are flat direction of both functions. Using eq. (\ref{admm}) we see
that the same property holds for the ADM mass: $\frac{\partial
}{\partial\phi^\alpha }M_{ADM}=0$.
Let us now give a general characterization of the $W$-function in
terms of the central and matter charges.
We can write the coset representative $\mathbb{L}(\phi^r)$ of ${\Scr
M}_{scal}$ as the product of the $G_0/H_0$ coset representative
$\mathbb{L}_0(\phi^\alpha)$ times a matrix ${\mathbb{L}_1}(\phi^k)$
depending on the remaining scalars:
\begin{eqnarray}
\mathbb{L}(\phi^r)&=&\mathbb{L}(\phi^\alpha,\phi^k)=
\mathbb{L}_0(\phi^\alpha)\,{\mathbb{L}_1}({\phi^k})\,.\label{altil}
\end{eqnarray}
We can write $\mathbb{L}(\phi^r)$ as a $2\,n_V\times 2\,n_V$ matrix
$\mathbb{L}(\phi^r)^M{}_{\hat{N}}$, where $M$ is an index in the real
symplectic representation, while $\hat{N}$ spans a complex basis in
which the action of $H$ is block-diagonal. We can obtain
$\mathbb{L}(\phi^r)^M{}_{\hat{N}}$ from the coset representative in
the real symplectic representation $\mathbb{L}_{Sp}(\phi^r)^M{}_{N}$
using the Cayley matrix:
\begin{eqnarray}
\mathbb{L}(\phi^r)&=&\mathbb{L}_{Sp}(\phi^r)\,\mathcal{A}^\dagger\,\,\,\,\,\,\,\,\mbox{where}\,\,\,\,\,\,\,\,\,
\mathcal{A}\equiv \frac{1}{\sqrt{2}}\,\left(\matrix{\bfone &
\ii\,\bfone\cr \bfone & -\ii\,\bfone }\right)\,.
\end{eqnarray}
The central and matter charges $Z_{AB},\,Z_I$ of the theory can be
arranged, together with their complex conjugates, in a
$(2\,n_V)$-vector $Z_{\hat{M}}$ defined as follows:
\begin{eqnarray}
Z_{\hat{M}}(\phi^r,\mathcal{P})&=&\left(\matrix{Z_{AB}\cr Z_I\cr
\bar{Z}^{AB}\cr
 \bar{Z}^I}\right)=-\mathbb{L}(\phi^r)^T\,\mathbb{C}\,\mathcal{P}=-{\mathbb{L}_1}({\phi^k})^T\,
\mathbb{L}_0(\phi^\alpha)^T\,\mathbb{C}\,\mathcal{P}\,,\label{zdef}
\end{eqnarray}
Now we can use the property of $\mathbb{L}_0(\phi^\alpha)$ of being
an element of $G_0$ in the symplectic representation, so that
$\mathbb{L}_0^T\,\mathbb{C}\,\mathcal{P}=\mathbb{C}\,\mathbb{L}_0^{-1}\,\mathcal{P}=\mathbb{C}\,\mathcal{P}$
and write:
\begin{eqnarray}
Z_{\hat{M}}(\phi^\alpha,\phi^k,\mathcal{P})&=&-{\mathbb{L}_1}(\phi^k)^T\,\mathbb{C}\,\mathcal{P}=
Z_{\hat{M}}(0,\phi^k,\mathcal{P})\,,
\end{eqnarray}
that is the central and matter charges do not depend on $\phi^\alpha$ at
all: \begin{eqnarray} \frac{\partial}{\partial {\phi^\alpha}
}Z_{AB}&=&\frac{\partial}{\partial {\phi^\alpha} }Z_{I}=0\,.
\end{eqnarray}
Let us now describe the effect of a generic transformation $g_0$ in
$G_0$ on the central charges. From the general properties of coset
representatives we know that
$D(g_0)\,\mathbb{L}_0(\phi^\alpha)=\mathbb{L}_0(g_0\star\phi^\alpha)\,D(h_0)$,
$D(h_0)$ being a compensator in $H_0$ depending on $g_0$ and
$\phi^\alpha$. Now, using the property that $\phi^k$ transform in a
linear representation of $H_0$, we can describe the action of $g_0$
on a generic point $\phi$ as follows:
\begin{eqnarray}
D(g_0)\,\mathbb{L}(\phi^r)&=&D(g_0)\,\mathbb{L}_0(\phi^\alpha)\,{\mathbb{L}_1}(\phi^k)=
\mathbb{L}_0(g_0\star\phi^\alpha)\,D(h_0)\,{\mathbb{L}_1}(\phi^k)\,D(h_0)^{-1}\,D(h_0)=\nonumber\\
&=& \mathbb{L}_0(g_0\star\phi^\alpha)\,{\mathbb{L}_1}({
\phi^{\prime\,k}})\,D(h_0)=\mathbb{L}(g_0\star\phi^r)\,D(h_0)\,,\label{Ltra}
\end{eqnarray}
where ${\phi^{\prime\,k}}$ is the transformed of $\phi^k$ by
$h_0$, and $(g_0\star\phi^\alpha,{\phi^{\prime\,k}})$ define the
transformed $g_0\star \phi^r$ of $\phi^r$ by $g_0$. From (\ref{Ltra})
and the definition (\ref{zdef}) we derive the following property:
\begin{eqnarray}
\forall g_0\in G_0&:&Z_{\hat{M}}(g_0\star
\phi^r,\mathcal{P})=[D(h_0)^{-T}]_{\hat{M}}{}^{\hat{N}}\,Z_{\hat{N}}(
\phi^r,\mathcal{P})=h_0\star Z_{\hat{M}}(
\phi^r,\mathcal{P})\,,\label{ztra}
\end{eqnarray}
where, to simplify notations we have denoted by $h_0\star Z$ the
vector $[D(h_0)^{-T}]_{\hat{M}}{}^{\hat{N}}\,Z_{\hat{N}}$.  Now
consider the $W$ function as a function of $\phi^r$ and $\mathcal{P}$
through the central and matter charges $Z_{\hat{M}}$:
\begin{eqnarray}
W(\phi^r,\mathcal{P})&=&\widehat{W}[Z_{\hat{M}}(\phi^r,\mathcal{P})]\,.\label{WZ}
\end{eqnarray}
From the duality-invariance of $W$ it follows that, for any $g_0\in
G_0$ we have
\begin{equation}
W(\phi^r,\mathcal{P})=W(g_0\star\phi^r,D(g_0)\,\mathcal{P})=W(g_0\star\phi^r,\mathcal{P})\,.
\end{equation}
 Furthermore, using eq.s (\ref{ztra}), (\ref{WZ}) we find:
 \begin{eqnarray}
\widehat{W}[Z(\phi^r,\mathcal{P})]&=&W(\phi^r,\mathcal{P})=W(g_0\star\phi^r,\mathcal{P})=
\widehat{W}[Z(g_0\star \phi^r,\mathcal{P})]=\widehat{W}[h_0\star
Z_{\hat{N}}( \phi^r,\mathcal{P})]\,.
 \end{eqnarray}
The above equality holds for any $g_0\in G_0$ and thus for any $h_0\in
H_0$. We conclude from this that \emph{$W$ can be characterized, for
a given orbit of solutions, as an $H_0$-invariant function of the
central and matter charges}. This is consistent with what was found
in \cite{HJ,Bossard:2009we}. Let us stress once more that we have
started from a generic charge vector $\mathcal{P}$, so that the
definition of $G_0$, and thus of $H_0$, is charge dependent. We
could have started from a given $G_0$ inside $G$ and worked out the
representative $\mathcal{P}_0$ of the $G$-orbit having $G_0$ as
manifest little group. In this case, by construction, the $({\phi^\alpha}
,\,\phi^k)$ parametrization is charge-independent.
\begin{table}\label{tabl}\begin{center}
{\scriptsize \begin{tabular}{|c||c|c|c|c|c|c|}
  \hline
  $\mathcal{N}$ & $\frac{G}{H}$ & orbit & $\frac{G_0}{H_0}$ & ${\bf R}_0$ & ${\bf R}_1$  \\\hline\hline
    &   &  I & $\frac{{\rm E}_{6(2)}}{{\rm SU}(2)\times {\rm SU}(6)}$ & ${\bf (2,20)}$ & ${\bf (1,15)}+c.r.$ \\
  8 & $\frac{{\rm E}_{7(7)}}{{\rm SU}(8)}$ & III & $\frac{{\rm E}_{6(6)}}{{\rm USp}(8)}$  & ${\bf 42}$ & ${\bf 1}+{\bf 27}$ \\\hline
    &   & I & $\frac{{\rm SU}(4,2)}{{\rm S}[{\rm U}(4)\times{\rm U}(2)]}$& ${\bf (4,2)}_{-3}+c.r.$ & ${\bf (6,1)}_{+2}+{\bf (1,1)}_{-4}+c.r.$ \\
  6 & $\frac{{\rm SO}^*(12)}{{\rm U}(6)}$  & II  & $-$  &   &   \\
    &   & III  & $\frac{{\rm SU}^*(6)}{{\rm USp}(6)}$  & ${\bf 14}$ &$2\times {\bf 1}+{\bf 14}$ \\\hline
  5 & $\frac{{\rm SU}(5,1)}{{\rm U}(5)}$ & I & $\frac{{\rm SU}(2,1)}{{\rm U}(2)}$  &
  ${\bf 2}_{+3}+c.r.$ & $3\times {\bf 1}_{-2}+c.r.$ \\\hline
   &   & I & $\frac{{\rm SO}(4,n)}{{\rm SO}(4)\times{\rm SO}(n)}$ & ${\bf (4,n)}$ &
    $2\times [{\bf (1,1)}+ {\bf (1,n)}]$\\
  4 & $\matrix{\frac{{\rm SL}(2,\mathbb{R})}{{\rm SO}(2)}\times \cr \times\frac{{\rm SO}(6,n)}{{\rm SO}(6)\times {\rm SO}(n)}}$ & II &
  $\frac{{\rm SO}(6,n-2)}{{\rm SO}(6)\times{\rm SO}(n-2)}$ & ${\bf (6,n-2)}$ &
    $2\times [{\bf (1,1)}+{\bf (6,1)}]$ \\
   &   & III & ${\rm SO}(1,1)\times\frac{{\rm SO}(5,n-1)}{{\rm SO}(5)\times{\rm SO}(n-1)}$  & ${\bf (1,1)}+{\bf (5,n-1)}$ &
   $\matrix{2\times {\bf (1,1)}+{\bf (5,1)}+\cr +{\bf (1,n-1)}}$ \\\hline
   &   & I & $\frac{{\rm SU}(2,n)}{{\rm S}[{\rm U}(2)\times{\rm U}(n)]}$ & ${\bf (2,n)}_{n+2}+c.r.$ & ${\bf (1,n)}_{-2}+c.r.$ \\
  3 & $\frac{{\rm SU}(3,n)}{{\rm S}[{\rm U}(3)\times{\rm U}(n)]}$ & II & $\frac{{\rm SU}(3,n-1)}{{\rm S}[{\rm U}(3)\times{\rm U}(n-1)]}$ & ${\bf (3,n-1)}_{n+2}+c.r.$ & ${\bf (3,1)}_{1-n}+c.r.$ \\\hline
    &  & I & $-$ &   &   \\
     & $\frac{{\rm SU}(1,n+1)}{{\rm U}(n+1)}$ & II &
    $\frac{{\rm SU}(1,n)}{{\rm U}(n)}$ & ${\bf n}_{n+1}+c.r.$ & ${\bf 1}_{-n}+c.r.$  \\\cline{2-6}
    &  & I & $-$ &   &   \\
    & $\matrix{\frac{{\rm SL}(2,\mathbb{R})}{{\rm SO}(2)}\times\cr\times \frac{{\rm SO}(2,n+2)}{{\rm SO}(2)\times {\rm SO}(n+2)}}$ & II &
    $\frac{{\rm SO}(2,n)}{{\rm SO}(2)\times {\rm SO}(n)}$ & ${\bf (2,n)}$ & $2\times [{\bf (2,1)}+{\bf (1,1)}]$  \\
    &   & III & ${\rm SO}(1,1)\times\frac{{\rm SO}(1,n+1)}{{\rm SO}(n+1)}$ & ${\bf 1}+{\bf (n+1)}$ & $3\times {\bf 1}+{\bf (n+1)}$ \\\cline{2-6}
   &  & I & $-$ &   &     \\
   & $\frac{{\rm Sp}(6)}{{\rm U}(3)}$ & II & $\frac{{\rm SU}(2,1)}{{\rm U}(2)}$ & ${\bf 2}_{-3}+c.r.$ & ${\bf 1}_{-4}+{\bf 3}_{+2}+c.r.$   \\
   &   & III & $\frac{{\rm SL}(3,\mathbb{R})}{{\rm SO}(3)}$ & ${\bf 5}$& $2\times{\bf 1}+{\bf 5}$   \\\cline{2-6}
    &   & I & $-$ & &  \\
2   & $\frac{{\rm SU}(3,3)}{{\rm S}[{\rm U}(3)\times{\rm U}(3)]}$ & II & $\left(\frac{{\rm SU}(2,1)}{{\rm U}(2)}\right)^2$ &
    ${\bf (2,1)}_{3,0}+{\bf (1,2)}_{0,3}+c.r.$ & ${\bf (2,2)}_{1,-1}+{\bf (1,1)}_{-2,2}+c.r.$  \\
    &   & III & $\frac{{\rm SL}(3,\mathbb{C})}{{\rm SU}(3)}$ & ${\bf 8}$ & $2\times {\bf 1}+{\bf 8}$   \\\cline{2-6}
    &   & I & $-$ &   &     \\
    & $\frac{{\rm SO}^*(12)}{{\rm U}(6)}$ & II &$\frac{{\rm SU}(4,2)}{{\rm S}[{\rm U}(4)\times{\rm U}(2)]}$& ${\bf (4,2)}_{-3}+c.r.$ & ${\bf (6,1)}_{+2}+{\bf (1,1)}_{-4}+c.r.$    \\
    &   & III & $\frac{{\rm SU}^*(6)}{{\rm USp}(6)}$  & ${\bf 14}$ &$2\times {\bf 1}+{\bf 14}$  \\\cline{2-6}
    &   & I & $-$ &   &     \\
    & $\frac{{\rm E}_{7(-25)}}{{\rm U}(1)\times {\rm E}_6}$ & II &$\frac{{\rm E}_{6(-14)}}{{\rm U}(1)\times {\rm SO}(10)}$&
    $\overline{{\bf 16}}_{+3}+c.r.$ & ${\bf 1}_{+4}+{\bf 10}_{-2}+c.r.$    \\
    &   & III & $\frac{{\rm E}_{6(-26)}}{{\rm F}_4}$  & ${\bf 26}$ &$2\times {\bf 1}+{\bf 26}$  \\
  \hline
\end{tabular}  }
\end{center}
 \caption{{\footnotesize  Summary of regular, extremal black hole orbits in the various supergravities. The symbols I, II, III denote the $\frac{1}{\mathcal{N}}$-BPS, the non-BPS ($I_4>0$) and the non-BPS ($I_4<0$) orbits respectively. For those solutions with
non-trivial moduli spaces $\frac{G_0}{H_0}$ (i.e. $G_0$ non-compact), the representations ${\bf R}_0,\,{\bf R}_1$ of $H_0$, see Appendix...., are given. The symbol ``c.r.'' stands for \emph{conjugate representations}.}}
\end{table}
\paragraph{A detailed analysis.}
Let us exploit now, for the BPS and non-BPS extremal, regular
solutions, the symmetry properties of the $W$ function discussed in
the previous sections, to study general aspects of the evolution of
the flat and non-flat directions.\par
 We start computing the Killing
vectors associated with the $G_0$-transformations and write the
condition that $W$ be $G_0$-invariant in the form of differential
equations. To this aim, we will first compute the general expression for the
vielbein of ${\Scr M}_{scal}$ in the parametrization (\ref{altil}).
Let us denote by $\{T_{\mathcal{A}}\}$, $\mathcal{A}=1,\dots, {\rm
dim}(G_0)$, the generators of $G_0$. We can perform the  Cartan
decomposition of the Lie algebras $\mathfrak{g}$ and
$\mathfrak{g}_0$ generating $G$ and $G_0$, respectively, with
respect to their maximal compact subalgebras
$\mathfrak{h},\,\mathfrak{h}_0$:
\begin{eqnarray}
\mathfrak{g}&=&\mathfrak{K}\oplus \mathfrak{h}\,\,\,,\,\,\,\,\,\,
\mathfrak{g}_0=\mathfrak{K}_0\oplus \mathfrak{h}_0\,.
\end{eqnarray}
Under the adjoint action of $H_0$, the space $\mathfrak{K}$ split
into subspaces $\mathfrak{K}_0$ and ${\mathfrak{K}}_1$ transforming
in the representations ${\bf R}_0$, ${{\bf R}_1}$ of $H_0$. The
non-compact generators $\{{\bf K}_{\hat{r}}\}$, $\hat{r}=1\,\dots,
n$, of $\mathfrak{K}$ (the indices $\hat{r},\,\hat{s}$ label basis
elements of the tangent space to the manifold) split into the
generators $\{{\bf K}_{a}\}$, $a,b=1,\dots, n_f$, of
$\mathfrak{K}_0$, belonging to the tangent space of the submanifold
$G_0/H_0$,  and the remaining $n-n_f$ generators $\{{\bf
K}_{\hat{k}}\}$ of ${\mathfrak{K}}_1$. The Lie algebra
$\mathfrak{h}_0$ of $H_0$ is generated by $\{{\bf H}_u\}$,
$u=1\,\dots, {\rm dim}(H_0)$. As far as the choice of the
parametrization is concerned, for the BPS and non-BPS ($I_4>0$)
solutions, we choose the coset representative as follows:
\begin{eqnarray}
\mathbb{L}(\phi^r)&=&\mathbb{L}_0(\phi^\alpha)\,{\mathbb{L}_1}(\phi^k)\in
e^{\mathfrak{K}_0}\cdot e^{{\mathfrak{K}}_1}\,,\label{l01}
\end{eqnarray}
that is $\mathbb{L}_0(\phi^\alpha)$ is an element of
$e^{\mathfrak{K}_0}\equiv G_0/H_0$ and ${\mathbb{L}_1}(\phi^k)$ is
an element of $e^{{\mathfrak{K}}_1}$. This in particular implies
that ${\phi^\alpha}$ and $\phi^k$ transform in the representations
${\bf R}_0$, ${{\bf R}_1}$ of $H_0$, respectively (see Table 1 for a
list of these representations).\par
 As for  the
non-BPS ($I_4<0$) solutions,  it is more convenient to adopt a
parametrization of the coset which is different from (\ref{l01}), in
which $\phi^k$ can be defined to transform linearly with respect to
the whole $G_0$. For this class of solutions, see below, we define
$\phi^k$ to be parameters of a solvable algebra
$\{s_k\}=\{s_\Lambda,\,s_0\}$, generated by $n-n_f-1$ nilpotent
generators $s_\Lambda$ and a Cartan generator $s_0$. As we shall
see, for the standard choice of the charges $\mathcal{P}_0$,
$\phi^k$ consist in $n-n_f-1$ axions originating from the $D=5$
vector fields and a dilaton describing the  modulus of the internal
radius in the $D=5\rightarrow D=4$ dimensional reduction.\par We
want to compute the components of the vielbein of ${\Scr M}_{scal}$
in the basis (\ref{l01}).
 To start with, the $G_0/H_0$ left-invariant 1-form
reads:
\begin{equation}
\Omega_0(\phi^\alpha)=L_0^{-1}dL_0=
\Omega_{0}^\mathcal{A}\,T_\mathcal{A}=d{\phi^\beta}\,
\Omega_{0\,\beta}{}^\mathcal{A}(\phi^\alpha)\,T_\mathcal{A}\,.
\end{equation}
If we split the $G_0$-generators $T_\mathcal A$ into generators of
$G_0/H_0$ (${\bf K}_{a}$) and of $H_0$ (${\bf H}_u$), that is
$\mathcal{A}\to (\tilde{\alpha},u)$, then
$\Omega_{0\,\beta}{}^{b}(\phi^\alpha)\equiv
V_\beta{}^{b}(\phi^\alpha)$ defines  the vielbein of $G_0/H_0$.
Moreover, let us introduce the left-invariant 1-form:
\begin{equation}
\Omega_1(\phi^k)\equiv
\mathbb{L}_1^{-1}d\mathbb{L}_1=d{\phi^k}\,V_k{}^{\hat{k}}(\phi^k){\bf
K}_{\hat{k}} + \mbox{connection}\,,\label{o1}
\end{equation}
where $d{\phi^k}\,V_k{}^{\hat{k}}$ will define the vielbein 1-forms
along the directions ${\bf K}_{\hat{k}}$ of the tangent space.
\footnote{The reason why the left-invariant 1-form $\Omega_1$ in eq.
(\ref{o1}) does not expand on the generators ${\bf K}_{a}$ will be
clarified in Appendix \ref{blockV}.}\par
 In terms of the above
quantities, we can now compute the left-invariant 1-form of ${\Scr
M}_{scal}$ in the basis (\ref{l01}):
\begin{eqnarray}
\Omega (\phi^r) &\equiv& \mathbb{L}^{-1}\,d\mathbb{L}=
{\mathbb{L}_1}^{-1}\,\Omega_0\,{\mathbb{L}_1}+{\mathbb{L}_1}^{-1}\,d{\mathbb{L}_1}=\Omega_0^{\mathcal{A}}(\phi^\alpha)\,
\mathbb{L}^{-1}_1 T_{\mathcal{A}}\mathbb{L}_1
+\Omega_1(\phi^k)\nonumber\\
&=&d{\phi^\beta}\,\Omega_{0\,\beta}{}^\mathcal{A}(\phi^\alpha)\,
{\mathbb{L}_1}_\mathcal{A}{}^{\hat{r}}(\phi^k)\,{\bf
K}_{\hat{r}}+d\phi^k\,V_k{}^{\hat{k}}\,{\bf
K}_{\hat{k}}+\mbox{connection}\,,
\end{eqnarray}
where we have written
${\mathbb{L}_1}^{-1}\,T_\mathcal{A}\,{\mathbb{L}_1}={\mathcal{
L}_1}_\mathcal{A}{}^{\hat{r}}(\phi^k)\,{\bf
K}_{\hat{r}}+\mbox{compact generators}$. Similarly we will also
write $\mathbb{L}_0^{-1}\,T_\mathcal{A}\,\mathbb{L}_0=\mathcal{
L}_{0\,\mathcal{A}}{}^{\mathcal{B}}(\phi^\alpha)\,T_\mathcal{B}$.
The non-vanishing components of the vielbein
$\mathcal{V}_r{}^{\hat{s}}$ of ${\Scr M}_{scal}$ are now readily
computed:
\begin{eqnarray}
\mathcal{V}_\beta{}^{b}&=&\Omega_{0\,\beta}{}^\mathcal{A}(\phi^\alpha)\,
{\mathcal{
L}_1}_\mathcal{A}{}^{b}(\phi^k)\,\,\,,\,\,\,\,\,\mathcal{V}_\beta{}^{\hat{k}}=\Omega_{0\,\beta}{}^\mathcal{A}(\phi^\alpha)\,
{\mathcal{
L}_1}_\mathcal{A}{}^{\hat{k}}(\phi^k)\,\,\,,\,\,\,\,\,\mathcal{V}_k{}^{\hat{k}}={V}_k{}^{\hat{k}}(\phi^k)\,.\label{vielbein}
\end{eqnarray}
Note that for all regular extremal black holes, our choice of
parametrization is such that the vielbein matrix has a vanishing
off-diagonal block ${V}_k{}^{a}$, see Appendix \ref{blockV}. \par
 The non-vanishing blocks of
the inverse vielbein $\mathcal{V}^{-1}{}_{\hat{r}}{}^r$ are:
\begin{equation}
\mathcal{V}^{-1}{}_{a}{}^\beta,\,
\mathcal{V}^{-1}{}_{\hat{k}}{}^k,\,\mathcal{V}^{-1}{}_{a}{}^k=-\mathcal{V}^{-1}{}_{a}{}^\beta\,\mathcal{V}_\beta{}^{\hat{k}}\,\mathcal{V}^{-1}{}_{\hat{k}}{}^k\,,\label{inverseviel}
\end{equation}
where
$\mathcal{V}^{-1}{}_{a}{}^\beta,\,\mathcal{V}^{-1}{}_{\hat{k}}{}^k$
are  the inverses of the diagonal blocks
$\mathcal{V}_\alpha{}^{b},\,\mathcal{V}_k{}^{\hat{k}}$,
respectively.\par Consider now an infinitesimal $G_0$-transformation
$g_0\sim {\bf 1}+\epsilon^\mathcal{A}\,T_\mathcal{A}$,
$\epsilon^\mathcal{A}\sim 0$, and write
$(g_0\star\phi)^r\sim\phi^r+\epsilon^\mathcal{A}\,k_\mathcal{A}^r(\phi)$.
The Killing vectors $k_\mathcal{A}^r(\phi)$ are computed, in the
parametrization (\ref{l01}), to be:
\begin{eqnarray}
k_\mathcal{A}^r&=& \mathcal{
L}_{0\,\mathcal{A}}{}^\mathcal{B}(\phi^\alpha)\,{\mathcal{
L}_1}_\mathcal{B}{}^{\hat{r}}(\phi^k)\,\mathcal{V}^{-1}{}_{\hat{r}}{}^r\,.\label{kill}
\end{eqnarray}
The $G_0$-invariance of the $W$-function ($W(g_0\star
\phi,\mathcal{P})=W(\phi,\mathcal{P})$) can now be expressed in the
following way:
\begin{eqnarray}
k_\mathcal{A}^r\,\frac{\partial W}{\partial
\phi^r}=0\,\,\,\Leftrightarrow \,\,\,\,\,\,{\mathcal{
L}_1}_\mathcal{A}{}^{\hat{r}}(\phi^k)\,\mathcal{V}^{-1}{}_{\hat{r}}{}^r\,\frac{\partial
W}{\partial \phi^r}=0\,,\label{kinvW}
\end{eqnarray}
where we have used the property that $\mathcal{
L}_{0\,\mathcal{A}}{}^\mathcal{B}(\alpha)$ is non-singular. Using
the expression of the vielbein, it will be useful to write the
first-order flow-equations for the scalar fields in the following
form:
\begin{eqnarray}
\dot{\phi}^r\,\mathcal{V}_r{}^{\hat{r}}&=&e^U\,\mathcal{V}^{-1\,\hat{r}\,s}\,\frac{\partial
W}{\partial
\phi^s}\,\,\,\Leftrightarrow\,\,\,\,\,\cases{\dot{\phi}^\beta\,\mathcal{V}_\beta{}^{a}=e^U\,\mathcal{V}^{-1\,a\,r}
\,\frac{\partial W}{\partial \phi^r}\cr
\dot{\phi}^r\,\mathcal{V}_r{}^{\hat{k}}=e^U\,\mathcal{V}^{-1\,\hat{k}\,k}
\,\frac{\partial W}{\partial \phi^k}}\,.\label{phiflo}
\end{eqnarray}
 We shall illustrate the
implications of the above formula in two relevant cases: The BPS
solution and the non-BPS one with $I_4<0$.
\paragraph{The BPS black holes.} For the sake of concreteness we
shall consider the supersymmetric regular solutions
($\frac{1}{8}$-BPS) in the maximal theory $\mathcal{N}=8$, although
our discussion is easily extended to non-maximal theories. In this
case $G_0=\rm E_{6(+2)}$ and $H_0=\rm SU(2)\times SU(6)\subset \rm
SU(8)=H$. With respect to the adjoint action of $H_0$, the coset
space $\mathfrak{K}$, in the ${\bf 70}$ of $\rm SU(8)$, splits into
the subspaces $\mathfrak{K}_0=\{{\bf K}_{a}\}$ in the ${\bf (2,20)}$
and ${\mathfrak{K}}_1=\{{\bf K}_{\hat{k}}\}$ in the ${\bf
(1,15)}\oplus {\bf (1,\bar{15})}$ of $H_0$, according to the
branching:
\begin{eqnarray}
{\bf 70}&\rightarrow &{\bf (2,20)}\oplus {\bf (1,15)}\oplus {\bf
(1,\bar{15})}\,.
\end{eqnarray}
The parametrization (\ref{altil}) amounts to the following choice of
the coset representative:
\begin{eqnarray}
\mathbb{L}&=&\mathbb{L}_0(\phi^\alpha)\,{\mathbb{L}_1}(\phi^k)\,\,\,,\,\,\,\,\,\mathbb{L}_0(\phi^\alpha)\in
e^{\mathfrak{K}_0}\,\,\,,\,\,\,\,\,{\mathbb{L}_1}(\phi^k)\in
e^{{\mathfrak{K}}_1}\,.
\end{eqnarray}
 Since the index $a$ spans a $\rm SU(2)$-doublet ($a=(A,\lambda)$, $A=1,2$, $\lambda=[mnp]=1,\,\dots,
 20$, $m,n,p=1,\dots, 6$),
while $\hat{k}$ only $\rm SU(2)$-singlets, being $\phi^k$ themselves
$\rm SU(2)$-singlets, the non vanishing components of the matrix
${\mathcal{ L}_1}_\mathcal{A}{}^{\hat{r}}$ are: ${\mathcal{
L}_1}_{a}{}^{b}(\phi^k),\, {\mathcal{ L}_1}_u{}^{\hat{k}}(\phi^k)$.
Consider now the implications of the $G_0$-invariance of $W$, as
expressed by eq. (\ref{kinvW}). The $H_0=\rm SU(2)\times
SU(6)$-invariance corresponds to the $\mathcal{A}=u$ component of
the equation, and implies
\begin{eqnarray}
{\mathcal{
L}_1}_u{}^{\hat{k}}(\phi^k)\,\mathcal{V}^{-1}{}_{\hat{k}}{}^k(\phi^k)\,\frac{\partial
W}{\partial \phi^k}=0\,.\label{hoW}
\end{eqnarray}
The invariance of $W$ under $G_0/H_0$-transformations, on the other
hand, implies, using (\ref{inverseviel}) and (\ref{vielbein}):
\begin{eqnarray}
0&=&{\mathcal{L}_1}_{a}{}^{b}(\phi^k)\,\mathcal{V}^{-1}{}_{b}{}^r\,\frac{\partial
W}{\partial
{\phi}^r}={\mathcal{L}_1}_{a}{}^{b}(\phi^k)\,\left[\mathcal{V}^{-1}{}_{b}{}^{\gamma}\,\frac{\partial
W}{\partial
\phi^\gamma}+\mathcal{V}^{-1}{}_{b}{}^{k}\,\frac{\partial
W}{\partial \phi^k}\right]=\nonumber\\&=&
{\mathcal{L}_1}_{a}{}^{b}(\phi^k)\,\left[\mathcal{V}^{-1}{}_{b}{}^{\gamma}\,\frac{\partial
W}{\partial
\phi^\gamma}-\mathcal{V}^{-1}{}_{b}{}^{\gamma}\,\Omega_{0\,\gamma}{}^u\,{\mathcal{L}_1}_u{}^{\hat{k}}\,\mathcal{V}^{-1}{}_{\hat{k}}{}^k\,\frac{\partial
W}{\partial \phi^k}\right]=
{\mathcal{L}_1}_{a}{}^{b}(\phi^k)\,\mathcal{V}^{-1}{}_{b}{}^{\gamma}\,\frac{\partial
W}{\partial \phi^\gamma}\,\,\,\Rightarrow\,\,\,\,\,\,\frac{\partial
W}{\partial {\phi^\alpha}}=0\,,\nonumber\\&&\label{gohoW}
\end{eqnarray}
where we have used eq. (\ref{hoW}) and the property that the block
${\mathcal{L}_1}_{a}{}^{b}(\phi^k)$ is non-singular. The above
equation expresses the $\phi^\alpha$-independence of $W$, which we
had proven before in a different way. Finally, consider the
evolution of the $\phi^\alpha $-scalars as described in
(\ref{phiflo}). From equation (\ref{gohoW}) it follows that:
\begin{eqnarray}
\dot{\phi}^\beta\,\mathcal{V}_\beta{}^{a}=e^U\,\mathcal{V}^{-1\,a\,r}
\,\frac{\partial W}{\partial \phi^r}=0\,,
\end{eqnarray}
namely \emph{the flat directions ${\phi^\alpha}$ are constant along the
flow}. This is consistent with the $\mathcal{N}=2$ supersymmetry of
the solution, since the variation of the fermions $\lambda^{mnp}$
(the hyperinos in the $\mathcal{N}=2$ truncation, in the ${\bf 20}$
of $\rm SU(6)$) on the solution reads:
\begin{eqnarray}
\delta \lambda^{mnp}&\propto &
\dot{\phi}^\alpha\mathcal{V}_\alpha{}^{{A},\,mnp}\,\epsilon_A=0\,,
\end{eqnarray}
where, as usual, we have written $a=(A,\,mnp)$.\par As far as the
non-BPS black holes with $I_4>0$ are concerned, the analysis is
analogous to the BPS case illustrated above.
\paragraph{Non-BPS black holes with $I_4<0$.}
In this case the little group $G_0$ of the charge vector is the
duality group of the five-dimensional parent theory (for the
$\mathcal{N}=8$ case $G_0=\rm E_{6(6)}$), so that the flat
directions $({\phi^\alpha})$ spanning $G_0/H_0$ are the five-dimensional
scalar fields. We can use the solvable parametrization for ${\Scr
M}_{scal}$ by writing ${\Scr M}_{scal}=\exp(Solv)$, where $Solv$ the
solvable Lie algebra defined by the Iwasawa decomposition of $G$
with respect to $H$. Let moreover $Solv_0$ be the solvable Lie
algebra generating the submanifold spanned by the flat directions:
$G_0/H_0\equiv \exp(Solv_0)$.

In the solvable parametrization the moduli ${\phi^\alpha}$ are
parameters of the generators $s_\alpha$ of $Solv_0$\footnote{Note
that, in contrast to the parametrization used for the other classes
of black holes, neither $\phi^\alpha$ nor the corresponding solvable
generators $s_\alpha$  transform, under the adjoint action of $H_0$,
in a linear representation.}.  We can decompose the scalars $\phi^r$
into ${\phi^\alpha}$ and $\phi^k$ by decomposing $Solv$ with respect
to $Solv_0$:
\begin{eqnarray}
Solv &=& \mathfrak{o}(1,1)\oplus Solv_0\oplus {{\bf
R}}_{-2}\,,\label{solvdec}
\end{eqnarray}
where the $\mathfrak{o}(1,1)$ generator $s_0$ is parametrized by the
modulus $\sigma_0$ of the radius of the fifth dimension and the
abelian subalgebra ${{\bf R}}_{-2}=\{s_\Lambda\}$ is parametrized by
the axions  $\sigma^\Lambda$ originating from the
five-dimensional vector fields and transforming according to the
representation $\bar{{\bf R}}$ of $G_0$ with $\rm O(1,1)$-grading
$+2$ (in the maximal theory ${\bf R}={\bf 27}$). The decomposition
(\ref{solvdec}) originates from the general branching rule of $G$
with respect to $G_0$
\begin{eqnarray}
{\rm Adj}(G) &=& {\bf 1}_0\oplus {\rm Adj}(G)\oplus {\bf
R}_{-2}\oplus \bar{{\bf R}}_{+2}\,,\label{gdec}
\end{eqnarray}
The non-flat directions $\phi^k$ therefore  consist of
$\sigma_0$ and $\sigma^\Lambda$, which transform in a
representation of $G_0$. The following commutation relations hold:
\begin{eqnarray}
[T_\mathcal{A},\,s_0]&=&0\,\,\,,\,\,\,\,[s_0,\,s_\Lambda]=+2\,s_\Lambda
\,\,\,,\,\,\,\,[T_\mathcal{A},\,s_\Lambda]=-T_{\mathcal{A}\,\Lambda}{}^\Sigma\,s_\Sigma\,.
\end{eqnarray}
We shall write
${\mathbb{L}_1}(\phi^k)={\mathbb{L}}(\sigma^\Lambda)\,e^{\sigma_0\,s_0}$.

Note that the coset parametrizations that we are using throughout
this section, defined in eq. (\ref{l01}), differ from the standard
parametrization of ${\Scr M}_{scal}$, which originates from the
$D=5\rightarrow D=4$ reduction (like, for instance, the special
coordinate parametrization of the special K\"ahler manifold in the
$\mathcal{N}=2$ theory).  The standard parametrization corresponds
indeed to the following choice of the coset representative:
\begin{eqnarray}
\mathbb{L}(\phi^r)&=&
\mathbb{L}(\tilde\sigma^\Lambda)\,e^{\sigma_0\,s_0}\,\mathbb{L}_0(\phi^\alpha)\,.
\end{eqnarray}
The prescription (\ref{altil}), that we are using here, yields
instead a different parametrization in which the order of the
factors in the coset representative is different:
$\mathbb{L}(\phi^r)=\mathbb{L}_0(\phi^\alpha)\,\mathbb{L}({\sigma}^{
\Lambda})\,e^{\sigma_0\,s_0}$. The two parametrizations are related
by a redefinition of the axions:
\begin{eqnarray}
{\tilde\sigma}^{
\Lambda}&=&\mathbb{L}^{-1}_0{}_{\Sigma}{}^\Lambda(\phi^\alpha)\,{\sigma}^\Sigma\,,\label{phiphip}
\end{eqnarray}
where $\mathbb{L}_{0\,\Sigma}{}^\Lambda(\phi^\alpha)$ is the matrix
form of $\mathbb{L}_0(\phi^\alpha)$ in  the ${\bf R}$
representation:
$\mathbb{L}_0(\phi^\alpha)^{-1}\,s_\Sigma\,\mathbb{L}_0(\phi^\alpha)=\mathbb{L}_{0\,\Sigma}{}^\Lambda(\phi^\alpha)\,s_\Lambda$.
The vielbein  1-forms $d\phi^r\,\mathcal{V}_{r}{}^{\hat{r}}$ are
defined, as usual, as the components of the left-invariant 1-form
along the non compact generators ${\bf K}_{\hat{k}}\propto
(s_r+s_r^\dagger)$. The non-vanishing components of the vielbein
matrix $\mathcal{V}_{r}{}^{\hat{r}}$ and of its inverse
$\mathcal{V}^{-1}{}_{\hat{r}}{}^{r}$ are readily computed to be:
\begin{eqnarray}
&&\mathcal{V}_\alpha{}^{b}(\phi^\alpha)\,\,\,,\,\,\,\,\mathcal{V}_\alpha{}^{\hat{\Lambda}}=-
e^{-2\,\sigma_0}\,V_\alpha{}^{b}(\phi^\alpha)\,s_{b\,\Sigma}{}^{\hat{\Lambda}}\,{\sigma}^\Sigma
\,\,\,,\,\,\,\,\mathcal{V}_\Lambda{}^{\hat{\Sigma}}=e^{-2\,\sigma_0}\,\delta_\Lambda{}^{\hat{\Sigma}}
\,\,\,,\,\,\,\,\mathcal{V}_0{}^{\hat{0}}=1\,,\nonumber\\
&&\mathcal{V}^{-1}{}_{a}{}^{\beta}(\phi^\alpha)\,\,\,,\,\,\,\,\mathcal{V}^{-1}{}_{a}{}^{\Lambda}=
s_{a\,\Sigma}{}^{{\Lambda}}\,\sigma^\Sigma\,\,\,,\,\,\,\,
\mathcal{V}^{-1}{}_{\hat{\Lambda}}{}^{{\Sigma}}=e^{2\,\sigma_0}\,\delta_{\hat{\Lambda}}{}^{{\Sigma}}
\,\,\,,\,\,\,\,\mathcal{V}^{-1}{}_{\hat{0}}{}^0=1\,,\label{explicitv}
\end{eqnarray}
where $s_{a\,\Sigma}{}^{{\Lambda}}$ is the matrix form of the
generator $s_{a}$ of $Solv_0$ in the representation ${\bf R}$.
Consider now the $G_0$-invariance condition on $W$, as expressed by
eq. (\ref{kinvW}) and use the following property:
\begin{eqnarray}
{\mathbb{L}_1}^{-1}\,T_\mathcal{A}\,{\mathbb{L}_1}&=&T_\mathcal{A}-
e^{-2\,\sigma_0}\,T_{\mathcal{A}\,\Sigma}{}^\Lambda\,\sigma^\Sigma=\mathcal{L}_{\mathcal{A}}{}^{\hat{r}}\,{\bf
K}_{\hat{r}}+\mbox{compact generators}\,.
\end{eqnarray}
After some algebra we find that the $H_0$-invariance of $W$
(component $\mathcal{A}=u$ of eq. (\ref{kinvW})) implies :
\begin{eqnarray}
H_{u\,\Sigma}{}^\Lambda\,\sigma^\Sigma\,\frac{\partial
W}{\partial \sigma^\Lambda}&=&0\,,\label{HWinv}
\end{eqnarray}
while the invariance with respect to $G_0/H_0$ (component
$\mathcal{A}=a$ of the same equation) implies:
\begin{eqnarray}
\frac{\partial W}{\partial {\phi^\alpha}}&=&0\,,
\end{eqnarray}
that is $W$ must be $\alpha$-independent, as expected by other
arguments. Let us note however that now the ${\phi^\alpha}$ are evolving
since:
\begin{eqnarray}
\dot{\phi}^\alpha\,\mathcal{V}_{\alpha\,a}&=&e^{U}\,s_{a\,\Sigma}{}^\Lambda\,\sigma^\Sigma\,
\frac{\partial W}{\partial \sigma^\Lambda}\neq 0\,,\label{adot}
\end{eqnarray}
since the right hand side represents the variation of $W$
corresponding to an infinitesimal $G_0/H_0$ transformation of
$\sigma^\Lambda$ and $W$ is invariant only with respect to
$H_0$-transformations of $\sigma^\Lambda$ (see equation
(\ref{HWinv})). One can easily verify that the flow of the non-flat
scalars $(\sigma_0,\,\sigma^\Lambda)$ is described by an
$\alpha$-independent dynamical system which has an equilibrium point
for  $\frac{\partial W}{\partial
\sigma^\Lambda}=\frac{\partial W}{\partial \sigma_0}=0$, at
which, by virtue of (\ref{adot}), also $\dot{\alpha}=0$. Indeed,
using eq.s (\ref{phiflo}) and  the explicit form of the vielbein
matrix and of its inverse (\ref{explicitv}), we can substitute in
the equations for $\phi^k$ the expression of
$\dot{\alpha}^a$ and find for the non-flat directions the following
equations:
\begin{eqnarray}
\dot\sigma^\Lambda&=&e^U\,\left(e^{4\,\sigma_0} \,
\delta^{\Lambda\Sigma}+s^{a}{}_\Delta{}^\Lambda\,s_{a}{}_\Gamma{}^\Sigma\,\sigma^\Delta\,\sigma^\Gamma\right)\,
\frac{\partial W}{\partial
\sigma^\Sigma}\,\,\,,\,\,\,\,\,\dot{\sigma}_0= e^U\,\frac{\partial
W}{\partial \sigma_0}\,.
\end{eqnarray}
According to the above equations, the non-flat directions
$\sigma^\Lambda,\,\sigma_0$ evolve towards fixed values at the
horizon which depend only on the quantized charges and solve the
equilibrium conditions $\frac{\partial W}{\partial
\sigma^\Lambda}=\frac{\partial W}{\partial \sigma_0}=0$. Only
the flat directions can depend at the horizon on the values of the
scalar fields at radial infinity, but this is not in contradiction
with the attractor mechanism since the near horizon geometry only
depends on the corresponding values of
$\sigma^\Lambda,\,\sigma_0$, through $V$ or $W$.
\par
 Let us finally give an example of  the
$({\phi^\alpha} ,\,\phi^k)$-parametrization in the STU model, in
the case $I_4(\mathcal{P})<0$, and show that the central and matter
charges do not depend on $\alpha$. The STU model is a
$\mathcal{N}=2$ supergravity with $n=6$ real scalar fields (i.e. $3$
complex ones $\{s,t,u\}\equiv \{z_1,\,z_2,\,z_3\}$) belonging to
three vector multiplets. The number of vector fields is $n_V=4$. The
scalar manifold has the following form:
\begin{eqnarray}
{\Scr M}_{STU}&=&\left(\frac{{\rm SL(2,\mathbb{R})}}{\rm
SO(2)}\right)_s\times \left(\frac{{\rm SL(2,\mathbb{R})}}{\rm
SO(2)}\right)_t\times \left(\frac{{\rm SL(2,\mathbb{R})}}{\rm
SO(2)}\right)_u\,,\label{mstu}
\end{eqnarray}
where each factor is parametrized by the complex scalars
$s=a_1'-\ii\,e^{\varphi_1},\,t=a_2'-\ii\,e^{\varphi_2},\,u=a_3'-\ii\,e^{\varphi_3}$.
The eight quantized charges transform in the ${\bf (2,2,2)}$ of the
isometry group $G={\rm SL(2,\mathbb{R})}^3$ and in this
representation the coset representative is the tensor product of the
coset representatives of each factor in (\ref{mstu}) in the
fundamental representation of ${\rm SL(2,\mathbb{R})}$:
\begin{eqnarray}
\mathbb{L}(z_i)&=&\mathbb{L}_1(z_1)\otimes \mathbb{L}_2(z_2)\otimes
\mathbb{L}_3(z_3)\,,
\end{eqnarray}
where each $2\times 2$ matrix has the following form:
\begin{eqnarray}
\mathbb{L}_i(z_i)&=&\left(\matrix{1 & 0\cr -a_i' &
1}\right)\,\left(\matrix{e^{-\frac{\varphi_i}{2}} & 0\cr 0&
e^{\frac{\varphi_i}{2}}}\right)\,.\label{li}
\end{eqnarray}
In this case the  ${\sigma}^{\prime\,\Lambda}$  axions are
nothing but $a_1',a_2',a_3'$. The little group of the
$I_4(\mathcal{P})<0$ orbit is $G_0={\rm O(1,1)^2}$. For generic
charges, like for instance those corresponding to the
$\overline{D0},\,D4$ system $(q_0,p^i)$, the action of $G_0$ is
rather involved and depends on the charges themselves. We can
consider however, as representative of the same $G$-orbit, the
charges corresponding to the $D0-D6$ system $(p^0,\,q_0)$. In this
case $G_0$ is parametrized by two combinations  of the dilatons
$\varphi_i$: 
$\{\phi^\alpha\}_{\alpha=1,2}=\{\phi^1=\frac{1}{\sqrt{2}}(\varphi_1-\varphi_2),\,\phi^2=\frac{1}{\sqrt{6}}(\varphi_1+\varphi_2-2\,\varphi_3)\}$.
According to the general prescription (\ref{altil}), the part
$\mathbb{L}_0$ of the coset representative  depending on the flat
directions $\phi^1,\,\phi^2$, should be the left factor of the
product. This corresponds to bringing the diagonal dilatonic factor
in (\ref{li}) to the left and redefining the axion:
\begin{eqnarray}
\mathbb{L}_i(z_i)&=&\left(\matrix{e^{-\frac{\varphi_i}{2}} & 0\cr 0&
e^{\frac{\varphi_i}{2}}}\right)\,\left(\matrix{1 & 0\cr -a_i &
1}\right)\,,\label{li2}
\end{eqnarray}
where $a_i=a_i'\,e^{-\varphi_i}$. The three complex scalar fields,
in this new parametrization, read: $z_i=e^{\varphi_i}\,(a_i-\ii)$.
The central and matter charges have the following form:
\begin{eqnarray}
Z&=&\frac{e^{\frac{\sqrt{3}}{2}\,\sigma_0}}{2\,\sqrt{2}}\,[q_0+p^0\,e^{\sqrt{3}\,\sigma_0}
(a_1-\ii)(a_2-\ii)(a_3-\ii)]\,,\\
Z_1&=&\frac{e^{\frac{\sqrt{3}}{2}\,\sigma_0}}{2\,\sqrt{2}}\,[q_0+p^0\,e^{\sqrt{3}\,\sigma_0}
(a_1+\ii)(a_2-\ii)(a_3-\ii)]\,,\\
Z_2&=&\frac{e^{\frac{\sqrt{3}}{2}\,\sigma_0}}{2\,\sqrt{2}}\,[q_0+p^0\,e^{\sqrt{3}\,\sigma_0}
(a_1-\ii)(a_2+\ii)(a_3-\ii)]\,,\\
Z_3&=&\frac{e^{\frac{\sqrt{3}}{2}\,\sigma_0}}{2\,\sqrt{2}}\,[q_0+p^0\,e^{\sqrt{3}\,\sigma_0}
(a_1-\ii)(a_2-\ii)(a_3+\ii)]\,,
\end{eqnarray}
where $\sigma_0\equiv
\frac{1}{\sqrt{3}}(\varphi_1+\varphi_2+\varphi_3)$. We observe that
none of the central and matter charges depend on the scalars
$\{\phi^\alpha\}=\{\phi^1,\,\phi^2\}$, but only on the remaining scalar fields
$\{\phi^k\}$, $k=3,\dots, 6$, defined as follows:
\begin{eqnarray}
\{\phi^k\}&=&\{\sigma_0,\,a_i\equiv a_i'\,e^{-\varphi_i}\}\,.
\end{eqnarray}
The scalars ${\phi^\alpha} $ are then flat directions of any function of
the central and matter charges, including $V$ and $W$.


\section{Small black holes in the $\mathcal{N}=8$ theory}
$\mathcal{N}=8$ supergravity admits two orbits for ``large''
extremal black holes (one of which is 1/4-BPS and the other a
non-BPS one) and three orbits for ``small'' extremal black holes
(all of them BPS, preserving 1/8, 1/4, and 1/2 supersymmetry
respectively).

Following the analysis of \cite{Cerchiai:2009pi}, the ADM mass for the three small orbits is given by the largest eigenvalue of the central charge matrix $Z_{AB}$.
Its eigenvalues for 1/8 and 1/4 BPS solutions are given by the quartic and quadratic roots of the secular equation
\begin{equation}\label{secular}
    \prod_{i=1}^{4} (\lambda -\lambda_i)=0  \qquad \quad \qquad (\lambda_i = \rho_i^2)  \,,
     \end{equation}
     $\rho_i$ being the skew-eigenvalue of $Z_{AB}$ when written in normal form.
     In particular we have:
     \begin{itemize}
       \item For 1/8 BPS: $\lambda_1>\lambda_2 \geq \lambda_3\geq\lambda_4$
        \item  For 1/4 BPS: $\lambda_1=\lambda_2 > \lambda_3=\lambda_4$
       \item For 1/2 BPS: $\lambda_i=\lambda$ $\forall$ $i=1,\cdots ,4$.
     \end{itemize}
     The five $\mathcal{N}=8$ orbits preserve, respectively, the following symmetries:

\begin{itemize}
\item large:
$\left\{
     \matrix{\mbox{ 1/8 BPS: \hskip 1cm }&\rm SU(2)\times SU(6)
      \cr \mbox{non-BPS: \hskip 1cm }&\rm USp(8) }
      \right.        $
\item small
 $\left\{   \matrix{\mbox{ 1/8 BPS: \hskip 1cm } &\rm USp(2)\times USp(6)\cr
      \mbox{ 1/4 BPS: \hskip 1cm } &\rm SU(4)\times USp(4)\cr
      \mbox{ 1/2 BPS: \hskip 1cm }&\rm  USp(8)}
      \right.   $
\end{itemize}
The superpotential $ W$, for all the BPS orbits, is given by the
highest eigenvalue of the central charge matrix $Z_{AB}$, however
one can also get small orbits from the large non-BPS orbit, in the
limiting procedure $I_4 \to 0$. Indeed, in this limit the non-BPS
orbit becomes supersymmetric, the fraction of supersymmetry
preserved depending on whether further constraints on $I_4$ are
imposed. For example, let us start with a non-BPS black hole with
charges $(p^0, q_0)$  turned on. It has $I_4= -(p_0q^0)^2$ and
symmetry $\rm USp(8)$. The limit $I_4 =0$, obtained for $ p_0q^0=0$,
gives a 1/2 BPS black home which has the same $\rm USp(8)$ symmetry.
For the most general ${W}$ of a non-BPS configuration, as defined in
\cite{Bossard:2009we}, the $I_4 =0$ limit just gives back eq.
(\ref{secular}) with $\lambda=W^2$.


\section{Small black holes in the $\mathcal{N}=4$ theory}
The moduli space of $\mathcal{N}=4$ supergravity is
$${\Scr M}_{scal} = \frac{\rm SO(1,2)}{\rm SO(2)}
\times \frac{\rm SO(6,n)}{\rm SO(6) \times SO(n)}\,.$$

The presence of a non-simple U-duality group $G=\rm SO(1,2)\times
SO(6,n)$ makes the analysis more involved here than in the
$\mathcal{N}=8$ case, and requires to explicitly set the notations.

We  consider static and spherically symmetric extremal black holes.
 In particular we address our study to the
small orbits of the theory, corresponding to a vanishing value of
the horizon area. They are identified by a vanishing quartic
invariant of the U-duality group.

The electric and magnetic charges span the representation
$\mathbf{(2,6+n)}$ of the U-duality group ${\rm SU(1,1)\times \rm
SO(6,n)}$\footnote{In this and in the following sections the indices
$r,a,\alpha,\Lambda$ have different range and definition with
respect to the previous sections. Their definition will be given as
soon as they are introduced.}:
$$\mathcal{P}^a_\Lambda\,,\qquad
a=1,2\,,\quad \Lambda =1,\cdots,6+n\,,$$ such that
\begin{equation}
\mathcal{P}^1_\Lambda = p_\Lambda=\eta_{\Lambda\Sigma}
p^\Sigma\,,\qquad \mathcal{P}^2_\Lambda = q_\Lambda \,.
\end{equation}
Here $\eta_{\Lambda\Sigma}=\mathrm {diag} (\buildrel
6\over{\overbrace{+,\cdots , +}},\buildrel
n\over{\overbrace{-,\cdots ,-}})$ is the
$\mathrm{SO(6,n)}$-invariant metric.

It is useful to introduce an $\rm SO(1,2)$-invariant tensor
quadratic in the charges:
$$ T_{\Lambda\Sigma}\equiv \frac 12 (p_\Lambda
q_\Sigma - q_\Lambda p_\Sigma),$$ and an $\rm SO(6,n)$-invariant
tensor, quadratic in the charges, in the adjoint representation of
$\mathrm{SO(1,2)}$, obtained as follows:
$$L^a = \frac 12 \gamma^a_{\alpha\beta}
\mathcal{P}^\alpha \cdot \mathcal{P}^\beta$$ where
$\gamma^a_{\alpha\beta}= (\bfone, -\sigma_3,
\sigma_1)_{\alpha\beta}$ and $\sigma^i$ denote the Pauli matrices.
The indices $a$ in the {\bf 3} of $\rm SO(1,2)$ are lowered and
raised with the metric $\eta_{ab}=\mathrm {diag} (+,-,-)$. The
explicit form of the 3-dimensional vector $L^a$ is given by:
\begin{equation}
L^0=\frac 12 (p^2+q^2)\,,\quad L^1 =\frac 12 (p^2-q^2)\,,\quad
L^2=p\cdot q
\end{equation}
where $p^2 \equiv p^\Lambda p_\Lambda$, $q^2 \equiv q^\Lambda
q_\Lambda$, $p\cdot q \equiv p^\Lambda q_\Lambda$.

The quartic U-invariant of the theory is given by
\begin{equation}I_4
=\epsilon_{\alpha\beta}\mathcal{P}^\alpha_\Lambda
\mathcal{P}^\beta_\Sigma \epsilon_{\gamma\delta}\mathcal{P}^{\gamma
\Lambda} \mathcal{P}^{\delta\Sigma}= p^2\,q^2 -(p\cdot q)^2 = L^a
\,L^b\,\eta_{ab}= 2 T_{\Lambda\Sigma}
T^{\Lambda\Sigma}\,.\label{quartic}
\end{equation}
The covariant tensors $T_{\Lambda\Sigma}$ and $L^a$ can be expressed
in terms of  derivatives of the quartic invariant, restricted to the
adjoint representation of the two subgroups of the U-duality group
as \cite{Cerchiai:2009pi}:
\begin{eqnarray}
T_{\Lambda\Sigma} &=&\frac
1{24}\epsilon^{\alpha\beta}\frac{\partial^2 I_4}{\partial
\mathcal{P}_\alpha^\Lambda
\partial \mathcal{P}_\beta^\Sigma }
=\left.\frac{\partial^2 I_4 }{\partial \mathcal{P}^2 }
\right\vert_{({\bf 1},{\bf Adj({\rm SO(6,n)})})} \\
L^a &=&\frac 1{8(5+n)} \gamma^{i|\alpha\beta}\frac{\partial^2
I_4}{\partial \mathcal{P}_\alpha^\Lambda
 \partial \mathcal{P}_{\beta\Lambda} } =
\left.\frac{\partial^2 I_4 }{\partial \mathcal{P}^2 }
 \right\vert_{({\bf Adj({\rm SO(1,2)})},{\bf 1})}
\end{eqnarray}
For the subsequent analysis we shall use the expression of $I_4$ in
terms of the central and matter charges given in (\ref{I44}). This
formula can be understood by noting that the vector
$$\tilde
L^a=(S_1,\mathrm{Re}(S_2),\mathrm{Im}(S_2))$$ transforms in the
$\mathbf{3}$ of $\rm SO(1,2)$, being related to $L^a$ through the
action of the coset representative of the $\frac {\rm SO(1,2)}{\rm
SO(2)}$ factor in the same representation. Therefore the 2 vectors
$\tilde L^a$ and $L^a$ are in the same duality orbit (indeed they
coincide in the origin of the $\frac {\rm SO(1,2)}{\rm SO(2)}$
factor). This is a relation between dressed and bare charges which
will be very useful in the sequel.\par
 When $I_4\geq 0$, the
sign of $L^0$ (and hence of $S_1$) has a U-duality invariant
meaning. Indeed, $I_4$ represents the norm of the vector $L^a$.
Using the terminology of the Lorentz group, a positive or null norm
vector
 $L^a$, being ``time-like'' or ``light-like''
 respectively, has
 the sign of its time component
 invariant under $\rm SO(1,2)$
 transformations. Viceversa, if $I_4<0$
 $L^a$ is space-like and the sign of its
 time-component has no invariant
 meaning.

 Exploiting the symmetries of
the theory, the central charge matrix can always be reduced to the
normal form $Z_{AB}\to \pmatrix{z_1 \epsilon & 0\cr 0& z_2
\epsilon}$ with skew eigenvalues $z_1,z_2 \in \mathbb{R}$, while the
matter charge-vector $Z_I =\rho_I e^{\ii \,\theta_I}$ can always be
reduced to a form where the first two components are $\rho_1\,
e^{\ii \theta}$ and $\rho_2$, all the other entries being zero
\cite{Cerchiai:2009pi,1order2}. In this normal frame $S_1$ and $S_2$
take the simple form:
\begin{eqnarray}
S_1&= &z_1^2 + z_2^2 -
\rho_1^2 -\rho_2^2\\
S_2 &= &2\,z_1\,z_2 -\rho_1^2\,e^{2\ii\,\theta} -\rho_2^2\,,
\end{eqnarray}
and the general expression for the quartic invariant in terms of
dressed charges in the normal frame reads:
\begin{eqnarray}
I_4&=& (z_1-z_2)^2 \left[\left(z_1+z_2\right)^2 -2\left(\rho_1^2
+\rho_2^2\right)\right]+2 \rho_1^2\,(\rho_2^2-2 z_1
z_2)\left(1-\cos(2\,\theta) \right)\,. \label{i40}
\end{eqnarray}
We shall however  use a different normal form for the matter
charges, in which central and matter charges appear in a more
symmetric fashion.
\begin{eqnarray}
\left(\matrix{Z_1^\prime\cr Z_2^\prime}\right)&\equiv
&\frac{1}{\sqrt{2}}\,\left(\matrix{1 & \ii\cr 1& -\ii
}\right)\,\left(\matrix{Z_1\cr Z_2}\right)=\frac{1}{\sqrt{2}}\,
\left(\matrix{\rho_1\,e^{\ii\,\theta}+\ii\,\rho_2\cr
\rho_1\,e^{\ii\,\theta}-\ii\,\rho_2}\right)=
e^{\ii\,\varphi}\,\left(\matrix{\widetilde{\rho}_1\,e^{\ii\,\beta}\cr
\widetilde{\rho}_2\,e^{-\ii\,\beta}}\right)\,.\label{nnf}
\end{eqnarray}
The phase $\beta$ can be absorbed by a ${\rm SO}(2)\subset {\rm
SO}(n)$ transformation which, in this new basis reads
$\left(\matrix{e^{-\ii\,\beta} & 0\cr 0& e^{\ii\,\beta} }\right)$.
Taking into account that
$(Z_1^\prime)^2+(Z_2^\prime)^2=2\,e^{2\,\ii\,\varphi}\,\widetilde{\rho}_1\,\widetilde{\rho}_2$,
$S_1$ and $S_2$ take the following form:
\begin{eqnarray}
S_1&=&z_1^2+z_2^2-\widetilde{\rho}_1^2-\widetilde{\rho}_2^2\,\,\,\,
\,\,\,\,\,S_2=2\,(z_1\,z_2-\widetilde{\rho}_1\,\widetilde{\rho}_2\,e^{2\,\ii\,\varphi})\,.
\end{eqnarray}
If we start from $Z_I=\rho_I\,e^{\ii\,\theta_I}$, $I=1,2$, and fix
$\theta_1-\theta_2=\pi/2$, using the ${\rm SO}(2)\subset {\rm
SO}(n)$ freedom, we easily find the following relations
$\widetilde{\rho}_1=\frac{\rho_1+\rho_2}{\sqrt{2}}$ and
$\widetilde{\rho}_2=\frac{\rho_1-\rho_2}{\sqrt{2}}$. The quartic
invariant in this new normal form reads:
\begin{eqnarray}
I_4&=&(\widetilde{\rho}_1^2-\widetilde{\rho}_2^2)^2+(z_1^2-z_2^2)^2-2\,(\widetilde{\rho}_1^2+\widetilde{\rho}_2^2)
\,(z_1^2+z_2^2)+8\,\widetilde{\rho}_1\,\widetilde{\rho}_2\,z_1\,z_2\,\cos(2\,\varphi)\,\label{I4}
\end{eqnarray}
Notice the symmetry between $z_1,z_2$ and
$\widetilde{\rho}_1,\,\widetilde{\rho}_2$. Indeed this normal form
can be easily obtained from the $\mathcal{N}=8$ central charges by
identifying $z_1,z_2,\,\widetilde{\rho}_1,\,\widetilde{\rho}_2$ with
the moduli $\rho_i$ of the skew-eigenvalues of $Z_{AB}$.\par
 It is straightforward to verify that for generic
$\widetilde{\rho}_1,\,\widetilde{\rho}_2$ the stabilizer on the
matter sector is ${\rm SO}(n-2)$. This does not change if either one
or the other of the two norms is put to zero. If, on the other hand,
$\widetilde{\rho}_1=\widetilde{\rho}_2$ the stabilizer is enhanced
to ${\rm SO}(n-1)$ and, finally, if the two matter charges are both
zero we recover the full ${\rm SO}(n)$. The five parameters
$\{z_1,\,z_2,\,\tilde{\rho}_1,\,\tilde{\rho}_2,\,\varphi\}$ defining
the normal form can be expressed in terms of the five $H$-invariant
functions of the central and matter charges characterizing a generic
configuration of scalar fields and charges at infinity, i.e. the
$H$-orbit of the solution. The classification of small and large
black holes, on the other hand, refers to the $G$-orbits of the
quantized charges $\mathcal{P}$. Each $G$-orbit will in general
comprise infinitely many $H$-orbits, defined by $G$-covariant
conditions on the five invariant parameters.
\subsection{Large orbits}
Let us first consider the  large black-hole solutions, with $I_4\neq
0$. We have three orbits, which can be characterized, all over the
flow, in terms of the sign of $I_4$ and $S_1$:
\begin{enumerate}
\item[$\alpha$)]
$1/4$-BPS orbit, preserving the symmetry $\rm SO(4)\times SO(n)$.

In this case $I_4>0$, and  $S_1>0$.
\item[$\beta$)]
non-BPS orbit with $Z_{AB}\neq 0$, preserving $\rm USp(4)\times
SO(n-1)$.

In this case $I_4 <0$,  and the sign of $S_1$ has no restrictions.
\item[$\gamma$)]
non-BPS orbit with $Z_{AB}=0$, preserving the symmetry $\rm
SU(4)\times SO(n-2)$.

In this case $I_4>0$, and  $S_1<0$.

\end{enumerate}

\subsection{Small orbits}
Let us now consider the case of small orbits, corresponding to
$I_4=0$. They can be classified, in terms of the invariants
introduced above, into 3 inequivalent classes
\cite{Cerchiai:2009pi}:
\begin{enumerate}
\item[A)] $T_{\Lambda\Sigma}=0$.

This class contains 3 different orbits, corresponding to $L^0$ being
positive, negative or null:
\begin{enumerate}
\item[A1)] $1/2$-BPS orbit preserving
the symmetry $\rm USp(4)\times SO(n)$.

In this case $L^0, S_1>0$.
\item[A2)] non-BPS orbit preserving
the symmetry $\rm SU(4)\times SO(n-1)$.

In this case  $L^0, S_1<0$.

\item[A3)] $1/2$-BPS orbit preserving
the symmetry $\rm USp(4)\times SO(n-1)$.

In this case  $L^\alpha=0,\, S_1=S_2=0$.
\end{enumerate}
We note that this class of orbits has a simple realization in the
heterotic basis, where the charges
$p_\Lambda=\mathcal{P}^1_\Lambda=0$. In this basis the three orbits
correspond to the norm of $q_\Lambda$ being positive, negative, or
null.
\item[B)]
$T_{\Lambda\Sigma}\neq 0$; $L^a=0$.

This is a $1/4$-BPS orbit preserving the symmetry $\rm SO(4)\times
SO(n-2)$.
\item[C)]$T_{\Lambda\Sigma}\neq 0$; $L^a\neq 0$.

It contains two orbits:
\begin{enumerate}
\item[C1)] $1/4$-BPS orbit preserving
the symmetry $\rm SO(4)\times SO(n-1)$.

In this case  $L^0, S_1>0$.
\item[C2)] non-BPS orbit preserving
the symmetry $\rm USp(4)\times SO(n-2)$.

In this case  $L^0, S_1<0$.
\end{enumerate}

\end{enumerate}

The above classification was found by studying the near-horizon
properties of the solutions. However, since the different orbits are
characterized in terms of U-duality invariants, actually the same
properties hold true all over the flow of the fields from space
infinity to the horizon.

\subsection{The $\mathcal{W}$ function for small orbits}
In the study of the large orbits of extremal black holes, the
dynamical flow can be completely characterized in terms of the
fake-superpotential $W$, which enjoys the property of being a
monotonic function decreasing from space-infinity (at $\tau =0$) and
the horizon (at $\tau\to -\infty$). In particular the ADM mass is
defined in terms of ${W}$ as:
\begin{equation}
M_{ADM}=\lim_{\tau \to 0} \dot U=\lim_{\tau \to 0} \frac 12
\mathcal{W(\phi(\tau))}=W(\phi_0)\,,
\end{equation}
and it is bounded from below by the value taken at the horizon,
which is fixed in terms of the charges
\begin{equation}
M_{ADM}=W(\phi_0)\geq {W}(\phi(-\infty))=
\sqrt{\frac{A_H}{4\,\pi}}\,.
\end{equation}
In this respect small orbits are problematic because in this case
the mass is not bounded from below. This raises the problem of
defining the fake superpotential for small orbits, where the
attractor mechanism breaks down. On the other hand, since $W$ is
well defined for the large orbits, we may nevertheless try to define
it  from the large orbit cases by an appropriate limiting procedure
obtained by choosing particular constraints equivalently on the bare
or on the dressed charges, such that the horizon area collapses to
zero: \begin{eqnarray} A_H=4\,\pi\,\sqrt{|I_4|}\to 0\,.
\end{eqnarray}
 This
requires a careful analysis of the behavior under this limit of the
invariants of the various orbits.
\subsection{Small versus large orbits}
In the following we give the constraints on the charges needed to
obtain zero-horizon area for each of the orbits listed above. To
this end let us describe  the large orbits in terms of the five
normal form parameters
$z_1,\,z_2,\,\tilde{\rho}_1,\,\tilde{\rho}_2,\,\varphi$. In
particular we shall consider representatives of the $G$-orbits of
the quantized charges and the corresponding values of the five
parameters on some specific point $\phi_0$ of the moduli space at
infinity, keeping in mind that on a generic point the solution is
characterized by five free parameters.
\begin{enumerate}
\item[$\alpha$)]{ Being the solution $1/4$-BPS,
\emph{in a generic point $\phi_0$ at infinity} the $W$ function is
given by the highest eigenvalue of the central charge matrix, say
$z_1$, expressed in terms of $H$-invariant functions of the central
and matter charges: $W=|z_1|$. As a representative of the orbit we
can take
 $\tilde{\rho}_2=0$ and denote $\tilde{\rho}_1=\rho$. In this case
$S_2$ is real and thus:
\begin{eqnarray}
I_4&=&
(S_1-S_2)\,(S_1+S_2)=[\rho^2-(z_1+z_2)^2]\,[\rho^2-(z_1-z_2)^2]\,.\label{alphI4}
\end{eqnarray}
Requiring it to be positive together with $S_1$ leads to the
condition $\rho^2<(z_1-z_2)^2$. As we shall see in the following, to
obtain the $A$-type small orbits by setting some of the charges to
zero, the above representative is not useful but a different one
should be chosen, with $\tilde{\rho}_1=\tilde{\rho}_2$.}
\item[$\beta$)]{We can choose $\tilde{\rho}_1=\tilde{\rho}_2=\rho$
and denote $z_1=z_2=z$. In this case we have:
\begin{eqnarray}
S_1&=&2\,(z^2-\rho^2)\,,\nonumber\\
S_2&=&2\,(z^2-\rho^2\,e^{2\,\ii\,\varphi})\,,\nonumber\\
I_4&=& -8\,\rho^2\,z^2\,(1-\cos(2\,\varphi)) <0\,.
\end{eqnarray}
 For
this orbit the general expression of $W$ in terms of $H$-invariant
quantities is not known;}
\item[$\gamma$)]{The $W$ function
for the $(\gamma)$ large orbit was computed in \cite{1order2}. There
it was shown that, fixing the relative phase of the two matter
charges to $\pi/2$, $W$ was simply expressed as
$W=\frac{1}{\sqrt{2}}\,(\rho_1+\rho_2)=\tilde{\rho}_1$, no
constraints being required on the invariant quantities
$z_1,\,z_2,\,\tilde{\rho}_I,\,\varphi$ characterizing the solution
at infinity. Just as in the $(\alpha)$ case, a representative of
this $G$-orbit can be chosen by setting
 $\tilde{\rho}_2=0$, $\tilde{\rho}_1$ being denoted by $\rho$.
   The quartic
 invariant is given by eq. (\ref{alphI4}) and the condition
 $I_4>0,\,S_1<0$ requires taking  $\rho^2>(z_1+z_2)^2$.\par This representative, however,
  is not useful to retrieve the $A$-type small orbits by simple constraints on the remaining
  charges and a different representative should be chosen (having $\tilde{\rho}_1=\tilde{\rho}_2$).
}
\end{enumerate}
 The large orbit $(\beta)$ has no constraints on $S_1$, so
it can generate all the small orbits by imposing appropriate
relations among the charges. However, this orbit is the only one
 for which we do not have a complete knowledge of
$W$, except for those $H$-orbits with fixed $\varphi$. We will then
be interested in deriving the small orbits from simple conditions on
the charges describing the $(\alpha)$ and $(\gamma)$ large ones, for
which $W$ is known. The $W$ functions of the small orbits will then be
obtained from those describing the $(\alpha)$- and
$(\gamma)$-solutions by imposing appropriate constraints on the
quantized charges.
\paragraph{The $A$-type orbits.}
As pointed out earlier, the  condition $T_{\Lambda\Sigma}=0$  can be
solved by setting $p_\Lambda=0$. Under this condition, as shown in
Appendix \ref{zeta}, central and matter charges satisfy a reality
condition, which, on the two normal forms, implies:
\begin{eqnarray}
\{z_1=z_2=z\,\,\,,\,\,\,\theta=0\}\,\,\,\Leftrightarrow\,\,\,\,\,
\{z_1=z_2=z\,\,\,,\,\,\,\tilde{\rho}_1=\tilde{\rho}_2=\tilde{\rho}\,\,\,,\,\,\,\varphi=0\}\,.\label{nfa}
\end{eqnarray}
We can easily see from (\ref{I4}) that the above conditions imply
$I_4=0$.
In particular, the condition $\varphi \to 0$ implies
\begin{equation}
I_4\to \left[(z_1-z_2)^2 -(\tilde{ \rho}_1-\tilde{\rho}_2)^2\right]
 \left[(z_1+z_2)^2 -(\tilde{ \rho}_1+\tilde{\rho}_2)^2\right]
 \label{i4phi0}
 \end{equation}
\begin{enumerate}
\item[A1)]
Since this orbit has $S_1>0$, it can be derived from the large
orbits $(\alpha)$ and  $(\beta)$. \par
The orbit $(\alpha)$  in the normal form has:
\begin{eqnarray}
z_1^2 +z_2^2&
>&\tilde\rho_1^2 +\tilde\rho_2^2\,.
\end{eqnarray}
 To obtain the $G$-orbit $(A1)$ from
the $(\alpha)$-orbit it suffices to set $z_1\rightarrow z_2$,
$\tilde{\rho}_1=\tilde{\rho}_2=0$ consistently with (\ref{nfa}).  Then from (\ref{i4phi0}) $I_4\to 0$.
The
compact symmetry of the charges is enhanced to $H_0=\rm
USp(4)\times SO(n)$ and the preserved amount of supersymmetry is
doubled to $1/2$-BPS.\par The $W$ function is given, in a generic
point on the moduli space at infinity, by the highest eigenvalue of
the central charge matrix: $W=z_1$;
\item[A2)]
It has $S_1<0$, so that it can be obtained from orbits $(\beta)$ and
$(\gamma)$. This is consistent with the fact that it is a non-BPS
orbit, which cannot be obtained from a BPS one by imposing relations
on the charges.
The orbit $(\gamma)$  in the normal form has:
\begin{eqnarray}
z_1^2 +z_2^2&
<&\tilde\rho_1^2 +\tilde\rho_2^2\,.
\end{eqnarray}
 To obtain a representative of the $G$-orbit $(A2)$ from
$(\gamma)$ we can start from a representative of the latter with
$z_1=z_2=0$ and impose  condition (\ref{nfa}) which requires setting
$\tilde{\rho}_1=\tilde{\rho}_2$.\par The compact stabilizer of the
charges is enlarged, with respect to the original orbit $(\gamma)$,
to $H_0=\rm SU(4)\times SO(n-1)$.\par The $W$ function is given, in
a generic point on the moduli space at infinity, by the highest of
the two matter charge moduli $\tilde{\rho}_I$: $W=\tilde{\rho}_1$;
\item[A3)]
This $A$-orbit has $S_1=0$. It can be obtained from  all the large orbits $(\alpha)$, $(\beta)$
and $(\gamma)$.
In particular, it is found from orbits $(\alpha)$ and $(\gamma)$,
 by imposing (\ref{nfa}), so that in particular $S_1$ reduces to
 \begin{equation}
 S_1= 2(z^2 -\tilde\rho^2)
 \end{equation} supplemented by the condition
\begin{equation}z^2=\tilde\rho^2\label{zrho}
\end{equation}
%
In this
limit the compact stabilizer of quantized charges reduces to
$H_0=\rm USp(4)\times SO(n-1)$.
%
%
Being the solution $1/2$-BPS, the $W$ function is given, in a
generic point on the moduli space at infinity, by the highest
eigenvalue of the central charge matrix: $W=z_1$.
 Note that eq. (\ref{zrho}) is a condition between central and matter charges, and it is a necessary condition,  when the orbit  is obtained from the $(\gamma)$ large orbit, for the enhancement of supersymmetry from non-BPS to 1/2-BPS, since for the $(\gamma)$ orbit $W$ is given by the highest eigenvalue of the matter charges, while in the BPS cases it is given by the highest eigenvalue of the central charges.

\end{enumerate}
\paragraph{The $B$ and $C$ type orbits.}  These orbits are
characterized by having $T_{\Lambda\Sigma}\neq 0$. This implies that
some of the conditions (\ref{nfa}) should be relaxed. In particular
we could have the following possibilities:
\begin{eqnarray}
B)&:& z_1\neq z_2\,\,\,,\,\,\,\,\tilde{\rho}_1\neq
\tilde{\rho}_2\,,\nonumber\\
C1)&:& z_1\neq z_2\,\,\,,\,\,\,\,\tilde{\rho}_1=
\tilde{\rho}_2\,,\nonumber\\
C2)&:& z_1= z_2\,\,\,,\,\,\,\,\tilde{\rho}_1\neq \tilde{\rho}_2\,.
\end{eqnarray}
Let examine these three cases in some detail.
\begin{enumerate}
\item[B)] The whole vector ${L_1}^a$ is zero in this case, which implies $S_1=S_2=0$.
 We can start from a representative of either the $(\alpha)$ or the $(\gamma)$-orbit
with $z_2=0=\tilde{\rho_2}$. In the former case
$z_1^2>\tilde{\rho}_1^2$ while in the latter
$z_1^2<\tilde{\rho}_1^2$, while $S_1$ and $I_4$ read:
\begin{eqnarray}
S_1&=&z_1^2-\tilde{\rho}_1^2\,\,\,,\,\,\,\,S_2=0\,,\nonumber\\
I_4&=&S_1^2=(z_1^2-\tilde{\rho}_1^2)^2\,.
\end{eqnarray}
To obtain a representative of the $B$-orbit we need to set:
\begin{equation}
z_1^2=\tilde{\rho}_1^2\label{zrhob}
\end{equation}
 The compact little group of this orbit is
$H_0=\rm SO(4)\times SO(n-2)$, and, being the solution $1/4$--BPS,
the $W$ function on a generic point of the moduli space is given by
the highest eigenvalue of the central charge matrix, say $W=z_1$.
Similarly to the $(A3)$ case, we note that the condition (\ref{zrhob})  between central and matter charges is necessary,  when the orbit  is obtained from the $(\gamma)$ large orbit, for the enhancement of supersymmetry from non-BPS to 1/4-BPS.

\item[C1)]
It has $S_1>0$ so it can be obtained from  orbits $(\alpha)$,
$(\beta)$. We can start from a representative of $(\alpha)$ with
$\varphi=0$ and $\tilde{\rho}_1=\tilde\rho_2=\tilde{\rho}$ (but $z_1> z_2$). In this case
 \begin{eqnarray}
I_4&=&(z_1-z_2)^2\,[(z_1+z_2)^2-4\,\tilde{\rho}^2]\ge
0\,.\end{eqnarray} If we further impose
$4\,\tilde{\rho}^2=(z_1+z_2)^2$, we find $I_4=0$ and $S_1=\frac 12\,
(z_1-z_2)^2\ge 0$. The compact little group of this $G$-orbit is
$H_0=\rm SO(4)\times SO(n-1)$ and, being it $1/4$-BPS, the $W$
function is given, in a generic point on the moduli space at
infinity, by the highest eigenvalue of the central charge matrix:
$W=z_1$.
\item[C2)]
It has $S_1<0$ so it can be obtained from the orbits $(\beta)$,
$(\gamma)$. In the latter case we can start from a representative
with $z_1=z_2=z$, $\tilde{\rho}_2=0$ and
$\tilde{\rho}=\tilde{\rho}_1\ge 4\,z^2$. In this case $S_1$ and
$I_4$ read:
\begin{eqnarray}
S_1&=&2\,z^2-\tilde{\rho}^2\le
0\,\,\,\,,\,\,\,\,\,I_4=\tilde{\rho}^2\,(\tilde{\rho}^2-4\,z^2)>0\,.
\end{eqnarray}
A representative of $C2$ is obtained by setting
$\tilde{\rho}^2=4\,z^2$.\par On a generic point in the moduli space,
the $W$ function is given by the highest of the two matter charge
moduli $\tilde{\rho}_I$, say $W=\tilde{\rho}_1$. \end{enumerate}

\section{Small Black Holes in $\mathcal{N}=2$ Magic Models}
We are going to show that, as in the $\mathcal{N}=4$ case, the
$\mathcal{N}=2$ ADM mass for small black holes either is
supersymmetric (and then given by $|Z(\phi_0)|$, in terms of the
asymptotic value at $\tau\to 0$ of the central charge) or can be
obtained by a non-BPS black hole with $Z=0$, and it has therefore a
known expression in terms of radicals
\cite{Andrianopoli:2007kz,Ceresole:2009vp}. An exceptional case in
this respect is the $t^3$-model which has no $Z=0$ orbit. In this
case, as we shall show below, the $W$ for the small orbit can be
obtained either from that of the BPS $I_4>0$ orbit or from the
$W$-function of the non-BPS $I_4<0$ orbit, which is known
\cite{Ceresole:2009iy}.
\par
\subsection{$D=6$ Uplift of Magic Models}
In $D=6$ there are three types of black holes, corresponding to the sign of the quadratic form
\begin{equation}
X^2 = X^A \eta_{AB} X^B\,,
\end{equation}
written in terms of the scalars in the tensor multiplets $X^A \in
\rm SO(q+1,1)/SO(q+1)$ ($q=1,2,4,8$ for real, complex, quaternionic
o octonionic models respectively). In particular, for $X^2 \neq  0$
we have large black holes, associated to two-charge solutions, BPS
for $X^2>0$, generated by two quantized charges of the same sign,
and non-BPS for $X^2<0$, corresponding to quantized charges of
opposite sign. For $X^2=0$ we have instead a small black hole, which
is a BPS one-charge solution.

The compact symmetry of the three different orbits are
\begin{itemize}
\item ${\rm SO(q+1)}$ \hskip 1cm for the BPS large black hole
\item ${\rm SO(q)}$ \hskip 1cm for the non-BPS large black hole
\item ${\rm SO(q)}$ \hskip 1cm for the BPS small black hole
\end{itemize}
We are going to consider in the following the octonionic case
($q=8$).
\subsection{$D=5$ Uplift}
There are three orbits of small black holes at $D=5$, corresponding
to the vanishing of the cubic invariant $I_3=0$. Two of them are
two-charge configurations, BPS if the 2 charges have the same sign,
non-BPS otherwise, while the third is a BPS one-charge
configuration, corresponding to $I_3=\partial I_3=0$. These three
orbits are the trivial dimensional reduction  of the six-dimensional
orbits. They preserve the compact symmetries ${\rm SO}(9)$, ${\rm
SO}(8)$ and ${\rm SO}(9)$  respectively.

Moreover, there are two large orbits corresponding to three-charge
configurations (with $I_3\neq 0$), a BPS one for charges of the same
sign, and a non-BPS one, when one of the charges has opposite sign
with respect to the other two. The compact symmetries preserved are
$F_4$ and $\rm SO(9)$ respectively.

\subsection{$D=4$ Analysis}
At $D=4$ there are five small orbits. Three of them originate from
dimensional reduction of the small $D=5$ orbits, and have compact
symmetry $\rm O(10)$, $\rm O(9)\times O(2)$ (critical, two charges)
and $\rm F_4$ (double critical, one charge); two of them are
three-charge light-like orbits coming by direct dimensional
reduction of the large $D=5$ orbits and have compact symmetries $\rm
F_4$ and $\rm SO(9)$.
\subsection{ADM Mass for Small Black Holes at $D=4$.}
For the three BPS orbits it is given by the asymptotic value, at
space infinity ($\tau\to 0$) of the norm of the central charge $Z$,
at $I_4=0$. For the two non-BPS orbits, both can be obtained by a
large non-BPS $Z=0$ orbit (with charges $(+,+,-,-)$) by setting one
of the charges to zero. Then the ADM mass is given by  the
asymptotic limit of $W$, which is obtained from the one of the
corresponding large orbit, which is a known radical function of the
H-invariants (see eq.s (5.11) and (5.12) of \cite{Ceresole:2009vp}
and eq. (3.4) of \cite{Bossard:2009we}).
\subsection{The $W$ Function of the Small Black Holes in the $t^3$--Model}
As anticipated, the $t^3$-model deserves a spacial treatment in our
analysis since it does not have the non-BPS orbit with $Z=0$.
However, in this case, both the $W$ functions for the $I_4>0$ (BPS)
and the $I_4<0$ (non-BPS) orbits are known. The former is
$W=\sqrt{i_1}=|Z|$, while the latter is given in eq. (3.15) of
\cite{Ceresole:2009iy} and reads:
\begin{eqnarray}
    \begin{array}{rcl}
        W^2&=&\displaystyle \frac{i_1+i_2}{4}+\frac{3}{8}\left[\left(\left(i_1-\frac{i_2}{3}\right)^3-(i_1+i_2)\,I_4+4\, i_3\,\sqrt{-I_4}\right)^{1/3}+ \right. \\[5mm]
        &&\displaystyle \left.+\left(\left(i_1-\frac{i_2}{3}\right)^3-(i_1+i_2)\,I_4- 4\, i_3\,\sqrt{-I_4}\right)^{1/3}\right].
    \end{array}
    \label{geom}
\end{eqnarray}
It is straightforward to verify that the above expression has a
finite $I_4\rightarrow 0$ limit which yields $W=\sqrt{i_1}=|Z|$. In
the $I_4\rightarrow 0$ limit therefore, both the $W$-functions
associated with the two large black hole orbits coincide with the
modulus of the central charge, as expected since small black holes
in this model are BPS.


 \appendix
\section{Proof of eq. (\ref{Wgen})}\label{proof}
Equation (\ref{Wgen}), as shown in
\cite{HJ}, is a particular form of the
general solution to the Hamilton-Jacobi
equation. In what follows we shall
tailor the formal proof given in
\cite{HJ} to the class of
 extremal solutions we are considering,
 without making use of the Hamilton-Jacobi formalism.
\par Consider the extremal solutions
$U(\tau;\phi_0,U_0)$ and
$\phi(\tau;\phi_0,U_0)$, for a given
charge vector $\mathcal{P}$, within the
interval $\tau_*<\tau<\tau_0$, where
now $U_0,\,\phi_0$ denote the values of
the fields computed at $\tau_0$:
$U_0=U(\tau_0;\phi_0,U_0),\,\phi_0=\phi(\tau_0;\phi_0,U_0)$.
The values of the fields at $\tau_*$,
for our family of solutions, is
completely fixed in terms of
$(U_0,\,\phi_0)$ and $\mathcal{P}$.
 Let us perform an
infinitesimal variation of the boundary conditions: $U_0\rightarrow
U_0+\delta U_0$ and $\phi_0=\phi_0+\delta\phi_0$. This will
determine a new solution within the same class:
\begin{eqnarray}
U(\tau;\phi_0+\delta\phi_0,U_0+\delta
U_0)&=&U(\tau;\phi_0,U_0)+\delta
U(\tau)\,,\nonumber\\\phi(\tau;\phi_0+\delta\phi_0,U_0+\delta
U_0)&=&\phi(\tau;\phi_0,U_0)+\delta \phi(\tau)\,.\label{varia}
\end{eqnarray}
Now we write a seemingly more general ansatz for $W$ than the one in
eq. (\ref{Wgen}):
\begin{eqnarray}
e^{U_0}\,W(\phi_0,\mathcal{P})&=&e^{U_*}\,W(\phi_*,\mathcal{P})+\int_{\tau_*}^{\tau_0}e^{2\,U(\tau;U_0,\phi_0)}\,V(\phi(\tau;U_0,\phi_0),\mathcal{P})\,d\tau\,.\label{Wgen2}
\end{eqnarray}
As we shall see, the result of this integral does not depend on the
choice of $\tau_0$. For the sake of simplicity we shall suppress the
dependence on $\tau$ and on the boundary values of the fields in the
integrand. Since the integral is computed along solutions, we can
use the Hamiltonian constraint (\ref{ham}) to rewrite $W$ as follows:
\begin{eqnarray}
e^{U_0}\,W(\phi_0,\mathcal{P})&=&e^{U_*}\,W(\phi_*,\mathcal{P})+\frac{1}{2}\,\int_{\tau_*}^{\tau_0}\left[
e^{2\,U}\,V(\phi,\mathcal{P})+\dot{U}^2+\frac{1}{2}\,G_{rs}\,\dot{\phi}^r\,\dot{\phi}^s
\right]\,d\tau=\nonumber\\&=&e^{U_*}\,W(\phi_*,\mathcal{P})+\frac{1}{2}\,\int_{\tau_*}^{\tau_0}\mathcal{L}_{eff}(U,\phi,\dot{U},\dot{\phi})\,d\tau\,.\nonumber
\end{eqnarray}
Now perform the variation (\ref{varia}), integrate by parts and use
the equations of motion:
\begin{eqnarray}
&&\delta U_0\,e^{U_0}\,W(\phi_0,\mathcal{P})+e^{U_0}\,\partial_r
W(\phi_0,\mathcal{P})\,\delta\phi^r_0=\delta(e^{U_*}\,W(\phi_*,\mathcal{P}))+\nonumber\\&&\quad+
\frac{1}{2}\,\int_{\tau_*}^{\tau_0}\left[\left(\frac{\partial}{\partial
U}\mathcal{L}_{eff}-\frac{d}{d\tau}\frac{\partial}{\partial
\dot{U}}\mathcal{L}_{eff}\right)\,\delta
U+\left(\frac{\partial}{\partial
\phi^r}\mathcal{L}_{eff}-\frac{d}{d\tau}\frac{\partial}{\partial
\dot{\phi}^r}\mathcal{L}_{eff}\right)\,\delta
\phi^r\right]+\nonumber\\&&\quad+\left.(\dot{U}\delta U+\frac{1}{2}
G_{rs}\,\dot{\phi}^s\,\delta
\phi^r)\right\vert_{\tau_*}^{\tau_0}=\delta(e^{U_*}\,W(\phi_*,\mathcal{P}))+\left.(\dot{U}\delta
U+\frac{1}{2} G_{rs}\,\dot{\phi}^s\,\delta
\phi^r)\right\vert_{\tau_*}^{\tau_0}\,,
\end{eqnarray}
where we have used the short-hand notation $\partial_r W\equiv
\frac{\partial W}{\partial \phi^r}$. We can choose $\tau_*=-\infty$,
so that all terms computed at $\tau_*$ in the above equation vanish.
Equating the variations at $\tau_0$ on both sides we find:
\begin{eqnarray}
\dot{U}(\tau_0)&=&e^{U_0}\,W(\phi_0,\mathcal{P})\,\,\,,\,\,\,\,\dot{\phi}^s(\tau_0)=2\,e^{U_0}\,G^{rs}(\phi_0)\,\partial_r
W(\phi_0,\mathcal{P})\,.
\end{eqnarray}
Being $\tau_0$ generic, we find that $W$ defines the first order
equations (\ref{1or}) for the fields and thus it is a solution to
the Hamilton-Jacobi equation.\par Note however that  $W$, as defined
in (\ref{Wgen2}), may in principle depend on the chosen value of
$\tau_0$, that is $W=W(U_0,\phi_0,\tau_0,\mathcal{P})$. Let us show
that this is not the case, namely that
$W(U_0,\phi_0,\tau_0+\delta\tau,\mathcal{P})=W(U_0,\phi_0,\tau_0,\mathcal{P})$,
for a generic $\delta \tau$. To do this we vary $\tau_0\rightarrow
\tau_0+\delta \tau$, \emph{keeping the boundary values of the fields
fixed}. This requires to change the solution on which the integral
is computed from $U(\tau),\phi(\tau)$ to
$U^\prime(\tau),\,\phi^\prime(\tau)$ such that:
\begin{eqnarray}
U^\prime(\tau_0+\delta\tau)&=&U(\tau_0)=U(\tau_0+\delta\tau)-\dot{U}(\tau_0)\,\delta
\tau\,,\nonumber\\\phi^\prime(\tau_0+\delta\tau)&=&\phi(\tau_0)=\phi(\tau_0+\delta\tau)-\dot{\phi}(\tau_0)\,\delta
\tau\,.
\end{eqnarray}
and thus  amounts to performing, along the flow, the transformation
$U\rightarrow U-\dot{U}\,\delta \tau,\,\phi\rightarrow
\phi-\dot{\phi}\,\delta \tau$, besides changing the domain of
integration, $\delta \tau$ being chosen along the flow so that
$\delta\tau_*=0$. After some straightforward calculations we find:
\begin{eqnarray}
e^{U_0}\,(W(U_0,\phi_0,\tau_0+\delta\tau,\mathcal{P})-W(U_0,\phi_0,\tau_0,\mathcal{P}))=\mathcal{H}_{eff}\vert_{
\tau_0}\,\delta \tau=0\,,
\end{eqnarray}
in virtue of the Hamiltonian constraint. Since the function $W$ of
the moduli space, as defined by (\ref{Wgen2}), does not depend on
the choice of $\tau_0$, we can choose $\tau_0=0$, where $U_0=0$ and
then find (\ref{Wgen}).
\section{Stability and Asymptotic Stability in the Sense of
Liapunov}\label{liapunov} Let us briefly recall the notion of
stability (in the sense of Liapunov) and of attractiveness of an
equilibrium point. Given an autonomous dynamical system:
\begin{eqnarray}
\dot{\phi}^r&=&f^r(\phi)\,,
\end{eqnarray}
 an
equilibrium point $\phi_*$ ($f^r(\phi_*)=0$),  is \emph{attractive}
(or an attractor), for $\tau\rightarrow -\infty$, if there exist a
neighborhood $\mathcal{I}_{\phi_*}$ of $\phi_*$, such that all
trajectories $\phi^r(\tau,\phi_0)$ originating at $\tau=0$ in
$\phi_0\in \mathcal{I}_{\phi_*}$ evolve towards $\phi_*$ as
$\tau\rightarrow -\infty$:
\begin{eqnarray}
\lim_{\tau\rightarrow -\infty}\phi^r(\tau,\phi_0)&=&
\phi_*^r\,\,\,,\,\,\,\,\forall \phi_0\in \mathcal{I}_{\phi_*}\,.
\end{eqnarray}
An equilibrium point  $\phi_*$ (not necessarily attractive) is
\emph{stable} (in the sense of Liapunov)  if, for any  $\epsilon>0$,
there exist a ball $\mathcal{B}_\delta(\phi_*)$ of radius $\delta>0$
centered in $\phi_*$, such that:
\begin{eqnarray}
\forall \phi_0\in \mathcal{B}_\delta(\phi_*)\,\,\,,\,\,\,\,\forall
\tau<0&: & \phi(\tau,\phi_0)\in \mathcal{B}_\epsilon(\phi_*)\,,
\end{eqnarray}
 that is, provided we take the starting point $\phi_0$ sufficiently close to $\phi_*$, the entire solution will
 stay, for all $\tau<0$, in any given, whatever small, neighborhood of $\phi_*$.
Finally an equilibrium point is \emph{asymptotically stable} (in
the sense of Liapunov) if it is attractive and stable.\par
\underline{\emph{Liapunov's Theorem}}: If there exist a function
$v(\phi)$ which is positive definite in a neighborhood of $\phi_*$
(that is positive in a neighborhood of $\phi_*$ and $v(\phi_*)=0$)
and such that also the derivative of $v$ along the solution, in
the same neighborhood, is positive definite\footnote{Here we
require positive definiteness because our critical point is
located at $\tau\rightarrow -\infty$ and not at $+\infty $ as in
standard textbooks.}: $\frac{dv}{d\tau}=\dot{\phi}^r\partial_r v>0
$, then $\phi_*$ is an asymptotically stable equilibrium point or,
equivalently, a stable attractor.
\par
For large extremal black holes such function is
$v(\phi)=W(\phi)-W(\phi_*)=W(\phi)-\sqrt{|I_4|}$.
\section{Properties of the Vielbein on $ {\Scr M}_{scal}$}
\label{blockV} Let us briefly motivate why, for all regular extremal
black holes, our choice of parametrization is such that the vielbein
matrix has a vanishing off-diagonal block ${V}_k{}^{a}$.\par
 The reason is purely group theoretical. As far as the BPS and non-BPS
($I_4>0$) orbits are concerned, taking into account that $\phi^k$
belong to ${{\bf R}_1}$ and the index $a$ label the ${\bf R}_0$
representation, $d\phi^k\,{V}_k{}^{a}(\phi^k)$ can be different from
zero only if ${\bf R}_0$ is contained in the tensor product of a
number of ${{\bf R}_1}$ representations. As the reader can ascertain
from Table 1, this is never the case.
 For example in the case of regular BPS black holes, for
$\mathcal{N}>2$, ${\bf R}_0$ is a doublet with respect to an ${\rm
SU}(2)$ subgroup of $H_0$, while ${{\bf R}_1}$ is a singlet with
respect to the same group. If we think of the $\mathcal{N}=2$
truncation of the original theory of which the same black hole is a
$1/2$-BPS solution, this ${\rm SU}(2)$ group is the quaternionic
structure of a quaternionic K\"ahler submanifold of the scalar
manifold spanned by the scalars ${\phi^\alpha}$. On the other hand,
as far as the non-BPS solutions with $I_4<0$ are concerned, the
above argument does not apply in the coset parametrization
(\ref{l01}), but choosing instead the solvable parametrization one
finds $d\phi^k{V}_k{}^{a}(\phi^k)=0$ since
${\mathbb{L}_1}^{-1}\,d{\mathbb{L}_1}$ belongs to the same solvable
algebra spanned by $\phi^k$, which is orthogonal to
$\mathfrak{K}_0$.
\section{An Explicit Parametrization of the $\mathcal{N}=4$ Scalar Manifold}\label{zeta}
The quantized charges $\mathcal{P}=(\mathcal{P}^M)$ of the
$\mathcal{N}=4$ model transform under $G=\rm SL(2,\mathbb{R})\times
SO(6,\,n)$ in the representation ${\bf (2,6+n)}$, which can be
labeled by the couple of indices $M=(\alpha,\Lambda)$, $\alpha=1,2$,
$\Lambda=1,\dots, 6+n$. In this basis the symplectic invariant
matrix reads:
\begin{eqnarray}
\mathbb{C}_{MN}&=&\mathbb{C}_{\alpha\Lambda,\,\beta
\Sigma}=-\epsilon_{\alpha\beta}\,\eta_{\Lambda\Sigma}\,,
\end{eqnarray}
Identifying the magnetic and electric charges with the components:
$\mathcal{P}^{1\,\Lambda}=p^\Lambda$, $\mathcal{P}^{2\,\Lambda}=q^\Lambda\equiv\eta^{\Lambda\Sigma}\,q_\Sigma$,
corresponds to choosing the
symplectic frame originating from the toroidal compactification of
the heterotic string, in which $\rm SO(6,n)$ acts block-diagonally.
We shall denote by $L_2=L_2{}^\Lambda{}_\Sigma$ the coset
representative of $\frac{\rm SO(6,n)}{\rm SO(6)\times SO(n)}$ and by
$L_1=L_1{}^\alpha{}_\beta$ that of $\frac{\rm
SL(2,\mathbb{R})}{\rm SO(2)}$, defined as:
\begin{eqnarray}
L_1{}^\alpha{}_\beta&=&\left(\matrix{e^{-\frac{\phi}{2}} & 0\cr -A\,
e^{-\frac{\phi}{2}} & e^{\frac{\phi}{2}}}\right)\,,\label{parasl2}
\end{eqnarray}
where $S=A-\ii\,e^{\phi}$ is the complex scalar parametrizing the
lower-half plane  $\frac{\rm SL(2,\mathbb{R})}{\rm SO(2)}$.\par
Using the definition (\ref{zdef}) of the central and matter charges,
we can write $Z_{\hat{M}}$ as $Z_{a\,\Lambda}$, where $a=1,2$ labels
the complex basis in which $\rm SO(2)$ acts diagonally and which is
obtained from the real one through the action of the Cayley matrix
$\mathcal{A}=(\mathcal{A}_{a}{}^\beta)=\frac{1}{\sqrt{2}}\,\left(\matrix{1
& \ii\cr 1& -\ii }\right)$. In this notation
$Z_{\hat{1}\,\Lambda}=(Z_r,\,Z_I)$, where
$Z_r=\frac{1}{2}\,\Gamma_{r}{}^{AB}\,Z_{AB}$, $r=1,\dots, 6$, are
the central charges and $Z_I$, $I=1,\dots, n$, are the matter ones.
We find:
\begin{eqnarray}
Z_{a\,\Lambda}&=&\mathcal{A}_{a}{}^\beta\,L_1{}^\sigma{}_\beta\,L_2{}^\Gamma{}_\Lambda\,
\epsilon_{\sigma \delta}\,\mathcal{P}^{\delta}_\Gamma\,,
\end{eqnarray}
 Using the explicit parametrization (\ref{parasl2}) we find:
\begin{eqnarray}
Z_{\hat{1}\,\Lambda}=(Z_r,\,Z_I)&=&\frac{1}{\sqrt{-2\,{\rm
Im}(S)}}\,L_2{}^\Gamma{}_\Lambda\,\left(q_\Gamma+S\,
p_\Gamma\right)\,.
\end{eqnarray}
We see that $(Z_r,\,Z_I)$ are real iff $p^\Lambda=0$. Reality of the
$Z_r$ implies the following reality condition on the $Z_{AB}$:
\begin{eqnarray}
Z_{AB}&=&\frac{1}{2}\,\epsilon_{ABCD}\,\overline{Z}^{CD}\,.
\end{eqnarray}
If $Z_{AB}$ is in the normal form, the above condition implies
$z_1=z_2$. Similarly, if the non vanishing matter charges
$Z_{I=1,2}=\rho_I$, characterizing the normal form, are real, from
(\ref{nnf}) we find that
$Z_1'=\frac{1}{\sqrt{2}}\,(\rho_1+\ii\,\rho_2)=\tilde{\rho}_1\,e^{\ii\,\beta},\,Z_2'=\overline{Z_1'}=
\tilde{\rho}_2\,e^{-\ii\,\beta}$, and thus that
$\tilde{\rho}_1=\tilde{\rho}_2$.
\section*{Acknowledgements}
The Work of L.A., R.D. and M.T. is supported in part by PRIN Program
2007 of MIUR and by INFN, sez. Torino. The work of S.F. is supported by ERC Advanced Grant
n.226455,``Supersymmetry,Quantum Gravity and Gauge
Fields''(Superfields), in part by  PRIN 2007-0240045 of Torino
Politecnico, in part by DOE Grant DE-FG03-91ER40662 and in part by INFN, sez. L.N.F.


\begin{thebibliography}{90}
\bibitem{1order}
A.~Ceresole and G.~Dall'Agata,
  ``Flow Equations for Non-BPS Extremal Black Holes,''
  JHEP {\bf 0703} (2007) 110
  [arXiv:hep-th/0702088].
 \bibitem{1order2}   L.~Andrianopoli, R.~D'Auria, E.~Orazi and M.~Trigiante,
  ``First Order Description of Black Holes in Moduli Space,''
  JHEP {\bf 0711} (2007) 032
  [arXiv:0706.0712 [hep-th]].
\bibitem{HJ}
  L.~Andrianopoli, R.~D'Auria, E.~Orazi and M.~Trigiante,
  ``First Order Description of D=4 static Black Holes and the Hamilton-Jacobi
  equation,''
  arXiv:0905.3938 [hep-th].
  \bibitem{Ceresole:2009iy}
  A.~Ceresole, G.~Dall'Agata, S.~Ferrara and A.~Yeranyan,
  ``First order flows for N=2 extremal black holes and duality invariants,''
  Nucl.\ Phys.\  B {\bf 824} (2010) 239
  [arXiv:0908.1110 [hep-th]].
  \bibitem{Bossard:2009we}
  G.~Bossard, Y.~Michel and B.~Pioline,
  ``Extremal black holes, nilpotent orbits and the true fake superpotential,''
  JHEP {\bf 1001} (2010) 038
  [arXiv:0908.1742 [hep-th]].
\bibitem{Ceresole:2009vp}
  A.~Ceresole, G.~Dall'Agata, S.~Ferrara and A.~Yeranyan,
  ``Universality of the superpotential for d = 4 extremal black holes,''
  arXiv:0910.2697 [hep-th].


\bibitem{fake} D.~Z.~Freedman, C.~Nunez, M.~Schnabl and K.~Skenderis,
  ``Fake Supergravity and Domain Wall Stability,''
  Phys.\ Rev.\  D {\bf 69} (2004) 104027
  [arXiv:hep-th/0312055].
\bibitem{Bellucci:2006xz}
  S.~Bellucci, S.~Ferrara, M.~Gunaydin and A.~Marrani,
  ``Charge orbits of symmetric special geometries and attractors,''
  Int.\ J.\ Mod.\ Phys.\  A {\bf 21} (2006) 5043
  [arXiv:hep-th/0606209];
 S.~Bellucci, S.~Ferrara, R.~Kallosh and A.~Marrani,
  ``Extremal Black Hole and Flux Vacua Attractors,''
  Lect.\ Notes Phys.\  {\bf 755} (2008) 115
  [arXiv:0711.4547 [hep-th]].
\bibitem{HJ2}V. I. Arnold,``Mathematical Methods of Classical Mechanics'' (Graduate Texts in Mathematics),
Springer, 1997; K. Meyer, G. Hall, ``Introduction to Hamiltonian
Dynamical Systems and the N-Body Problem,'' , Springer-Verlag, 1992.
\bibitem{DW}
 J.~de Boer, E.~P.~Verlinde and H.~L.~Verlinde,
        ``On the holographic renormalization group,''
        JHEP {\bf 0008} (2000) 003
        [arXiv:hep-th/9912012];
        E.~P.~Verlinde and H.~L.~Verlinde,
``RG-flow, gravity and the cosmological constant,''
  JHEP {\bf 0005} (2000) 034
  [arXiv:hep-th/9912018];
   M.~Fukuma, S.~Matsuura and T.~Sakai,
  ``Holographic renormalization group,''
  Prog.\ Theor.\ Phys.\  {\bf 109} (2003) 489
  [arXiv:hep-th/0212314];
    K.~Skenderis and P.~K.~Townsend,
  ``Hamilton--Jacobi method for Domain Walls and Cosmologies,''
  Phys.\ Rev.\  D {\bf 74} (2006) 125008
  [arXiv:hep-th/0609056];
  P.~K.~Townsend,
  ``Hamilton-Jacobi Mechanics from Pseudo-Supersymmetry,''
  Class.\ Quant.\ Grav.\  {\bf 25} (2008) 045017
  [arXiv:0710.5178 [hep-th]];
    K.~Hotta,
  ``Holographic RG flow dual to attractor flow in extremal black holes,''
  arXiv:0902.3529 [hep-th];
    B.~Janssen, P.~Smyth, T.~Van Riet and B.~Vercnocke,
  ``A first-order formalism for timelike and spacelike brane solutions,''
  JHEP {\bf 0804} (2008) 007
  [arXiv:0712.2808 [hep-th]].
  \bibitem{Liap}
W. Hahn, ``Stability of Motion,'' Springer-Verlag, 1967; N. Rouche
and J. Mawhin, ``Ordinary Differential Equations. Stability and
Periodic Solutions,'' Pitman, Boston/London/Melbourne, 1980.
 \bibitem{Ferrara:2007tu}
  S.~Ferrara and A.~Marrani,
 ``On the Moduli Space of non-BPS Attractors for N=2 Symmetric Manifolds,''
  Phys.\ Lett.\  B {\bf 652} (2007) 111
  [arXiv:0706.1667 [hep-th]].
\bibitem{bfk}
M.~Bianchi, S.~Ferrara and R.~Kallosh,
  ``Perturbative and Non-perturbative N =8 Supergravity,''
  arXiv:0910.3674 [hep-th];
M.~Bianchi, S.~Ferrara and R.~Kallosh,
  ``Observations on Arithmetic Invariants and U-Duality Orbits in N =8
  Supergravity,''
  arXiv:0912.0057 [hep-th].
\bibitem{Cerchiai:2009pi}
  B.~L.~Cerchiai, S.~Ferrara, A.~Marrani and B.~Zumino,
  ``Duality, Entropy and ADM Mass in Supergravity,''
  Phys.\ Rev.\  D {\bf 79} (2009) 125010
  [arXiv:0902.3973 [hep-th]].

\bibitem{Andrianopoli:2007kz}
L.~Andrianopoli, S.~Ferrara, A.~Marrani and M.~Trigiante,
  ``Non-BPS Attractors in 5d and 6d Extended Supergravity,''
  Nucl.\ Phys.\  B {\bf 795} (2008) 428
  [arXiv:0709.3488 [hep-th]].


  \bibitem{bhreviews}
  L.~Andrianopoli, R.~D'Auria, S.~Ferrara and M.~Trigiante,
  ``Extremal black holes in supergravity,''
  Lect.\ Notes Phys.\  {\bf 737} (2008) 661
  [arXiv:hep-th/0611345];
 P.~Aschieri, S.~Ferrara and B.~Zumino,
  ``Duality Rotations in Nonlinear Electrodynamics and in Extended
  Supergravity,''
  Riv.\ Nuovo Cim.\  {\bf 31} (2009) 625
  [Riv.\ Nuovo Cim.\  {\bf 031} (2008) 625]
  [arXiv:0807.4039 [hep-th]];
 S.~Bellucci, S.~Ferrara, M.~Gunaydin and A.~Marrani,
  ``SAM Lectures on Extremal Black Holes in d=4 Extended Supergravity,''
  arXiv:0905.3739 [hep-th].
  \bibitem{Gimon:2007mh}
  E.~G.~Gimon, F.~Larsen and J.~Simon,
  ``Black Holes in Supergravity: the non-BPS Branch,''
  JHEP {\bf 0801} (2008) 040
  [arXiv:0710.4967 [hep-th]].
  \bibitem{cj}
  E.~Cremmer and B.~Julia,
  ``The SO(8) Supergravity,''
  Nucl.\ Phys.\  B {\bf 159} (1979) 141.
 \bibitem{kalkol}
   R.~Kallosh and B.~Kol,
 ``E(7) Symmetric Area of the Black Hole Horizon,''
  Phys.\ Rev.\  D {\bf 53} (1996) 5344
  [arXiv:hep-th/9602014].
\end{thebibliography}
\end{document}